\documentclass[a4paper,12pt,fleqn]{article}
\usepackage[DIV13]{typearea}
\usepackage{amsmath}
\usepackage{amssymb}
\usepackage{amsfonts}
\usepackage{cite}
\usepackage{bbold}
\usepackage[footnotesize]{caption2}
\usepackage{graphicx}
\usepackage[center,footnotesize,hang]{subfigure}
\usepackage{url}

\newcommand{\Delatm}{\Delta m_{31} ^{2}}
\newcommand{\Delsol}{\Delta m_{21} ^{2}}
\newcommand{\eV}{\ensuremath{\,\mathrm{eV}}}

\newcommand{\MeV}{\ensuremath{\,\mathrm{MeV}}}
\newcommand{\GeV}{\ensuremath{\,\mathrm{GeV}}}
\newcommand{\TeV}{\ensuremath{\,\mathrm{TeV}}}

\newcommand{\Cl}[1]{\mathcal{C} _{#1}}

\newcommand{\Ord}[2]{\; ^{\circ} \mathrm{#1}_{#2}  \;}
\newcommand{\OrdCl}[1]{\; ^{\circ} \mathcal{C} _{#1} \;}
\newcommand{\Rep}[1]{\underline{\mbox{\textbf{#1}}}}
\newcommand{\MoreRep}[2]{\underline{\mbox{\textbf{#1}}} _{\mbox{\textbf{#2}}}}
\newcommand{\Groupname}[2]{$ {#1} _{#2} $}
\newcommand{\Doub}[2]{$ {#1} _{#2} ^{\prime} $}

\DeclareMathOperator{\diag}{diag}

\newcommand{\Eqref}[1]{Eq.\eqref{#1}}

\newcommand{\Tabref}[1]{Table \ref{#1}}

\newcommand{\Secref}[1]{Section \ref{#1}}
\newcommand{\Appref}[1]{Appendix \ref{#1}}

\begin{document}

\begin{titlepage}

\ \vspace*{-15mm}
\begin{flushright}
TUM-HEP-628/06
\end{flushright}
\vspace*{25mm}

\begin{center}
{\Large\sffamily\bfseries
The discrete flavor symmetry \mathversion{bold} \Groupname{D}{5} \mathversion{normal}}
\\[13mm]
{\large
C. Hagedorn\footnote{E-mail: \texttt{chagedor@ph.tum.de}}$^{\, a)}$,
M. Lindner\footnote{E-mail: \texttt{lindner@ph.tum.de}}$^{\, a),\, b)}$  and
F. Plentinger\footnote{E-mail: \texttt{fplentin@ph.tum.de}}}$^{\, a)}$ 
\\[5mm]
{\small $^{a)}$ \textit{
Physik-Department T30d, Technische Universit\"{a}t M\"{u}nchen\\
James-Franck-Stra{\ss}e, 85748 Garching, Germany
}}
\\[3mm]
\small $^{b)}$ \it{
Max-Planck-Institut f\"{u}r Kernphysik\\ 
Postfach 10 39 80, 69029 Heidelberg, Germany
}
\vspace*{1.0cm}
\end{center}
\normalsize
\begin{abstract}

\noindent We consider the standard model (SM) extended by the flavor symmetry
\Groupname{D}{5} and search for a minimal model leading to viable
phenomenology. We find
that it contains
four Higgs fields apart from the three generations of fermions whose
left- and left-handed conjugate parts do not transform in the same way under
\Groupname{D}{5}. We provide two numerical fits for the
case of Dirac
and Majorana neutrinos to show the
viability of our low energy model. The fits can accommodate all data
with the neutrinos being normally
ordered. For Majorana neutrinos two of the right-handed neutrinos are
degenerate. Concerning the Higgs sector we find that all potentials
constructed with three SM-like Higgs doublets transforming as
$\Rep{1}+\Rep{2}$ under \Groupname{D}{5} have a further unwanted global
$U(1)$ symmetry. Therefore we consider the case of four Higgs fields
forming two \Groupname{D}{5} doublets and show that this potential
leads to viable solutions in general, however it does not allow
spontaneous CP-violation (SCPV) for an arbitrary vacuum expectation
value (VEV) configuration. Finally, we discuss extensions of our model
to grand unified theories (GUTs) as well as embeddings of
\Groupname{D}{5} into the continuous flavor symmetries $SO(3)_{f}$
and $SU(3)_{f}$.

\end{abstract}

\end{titlepage}

\setcounter{footnote}{0}

\section{Introduction}
\label{sec:intro}

Gauge interactions and charge quantization of quarks and leptons
can successfully be described by the mathematical concept of Lie
groups, e.g. in the framework of GUTs. Albeit the number of fermion
generations, the diverse masses and mixing parameters of quarks and
leptons remain free parameters. It is tempting to assume that these
properties can also be explained by some (flavor) symmetry $G_{f}$. For several
reasons $G_{f}$ is chosen to be discrete and non-abelian in many
models. In the literature the permutation
symmetries \Groupname{S}{3} \cite{s3}, \Groupname{A}{4} \cite{a4} and
\Groupname{S}{4} \cite{s4}, the single- and double-valued dihedral
groups such as \Groupname{D}{4} \cite{d4} and
\Doub{D}{2} \cite{d2prime} and groups \Groupname{D}{n},
\Doub{D}{n} with larger index $n$ \cite{d5,dns,dnprimes} have been discussed. Furthermore the
two-valued group \Doub{T}{} \cite{tprime} and subgroups of $SU(3)$,
$\Delta (48)$ and $\Delta(75)$ \cite{deltas}, belonging to the series
of $\Delta (3 n^2)$ and $\Delta (6 n^2)$ with $n \in \mathbb{N}$ have
been studied. Most of
these groups have been used to maintain a certain fermion mass
texture. However, proceeding in this way does not answer the question
which fundamental group structure of a discrete symmetry is favorable
for describing nature and which is not. 

\noindent In order to investigate the generic features of a certain
group structure it is enough to discuss the
smallest group which reveals this structure. Therefore we choose the
flavor symmetry to be \Groupname{D}{5} which is the smallest group with two irreducible (faithful)
inequivalent two-dimensional representations. This group is used in \cite{d5} to produce 
certain mass textures for the lepton sector, but mass matrices for the
quarks as well as the Higgs sector are not discussed. Apart
from \Groupname{D}{5} only the discussed groups \Groupname{D}{n} for
$n \geq 6$, \Doub{D}{n} for $n > 2$ and \Doub{T}{} have more than one irreducible
two-dimensional representation. However, in general the groups differ
in the product structure. 

\noindent Our starting point is thus the SM gauge group extended by the
flavor group \Groupname{D}{5}. Both groups are broken only
spontaneously at the electroweak scale. We require a partial
unification for left- and left-handed conjugate fields, i.e. both
should transform as $\Rep{1} + \Rep{2}$ under \Groupname{D}{5} where
$\Rep{1}$ and $\Rep{2}$ do not need to be the same for both. Since we
do not want to give up the idea of unified gauge groups we
further require that our model is embeddable into the Pati-Salam
group $SU(4)_{C} \times SU(2)_{L} \times SU(2)_{R}$, $SU(5)$, $SO(10)$ or $E(6)$. The
resulting mass matrices should allow a viable fit of all data which
will be demonstrated by numerical examples. For this and for the spontaneous
breaking of \Groupname{D}{5}, we have to take at least three $SU(2)
_{L}$ doublet Higgs fields which transform non-trivially under \Groupname{D}{5}. Since
there exist strong bounds on flavor changing neutral currents (FCNCs),
the number of Higgs fields should be as small as possible and they
should be sufficiently heavy. Furthermore
we discard the possible existence of $SU(2) _{L}$ triplet and SM
gauge singlet (scalar) fields. Taking all these constraints and the
requirement that there are no left-over massless Goldstone bosons coming
from accidental symmetries of the Higgs potential we will show that we need
at least four Higgs fields. With these it turns out to be favorable to have
different transformation properties of left- and left-handed
conjugate fermions under \Groupname{D}{5}. The neutrinos can be
either Dirac or Majorana particles. In the second case two of the
right-handed neutrinos are degenerate, since there are no SM gauge
singlets in the theory. 

\noindent We contrast this minimal \Groupname{D}{5} invariant model
with the corresponding one invariant under the flavor symmetry \Groupname{D}{3}
which is isomorphic to \Groupname{S}{3} and considered very often in
the literature \cite{s3}. 

\noindent We also discuss the three Higgs potential in detail and show
the existence of an accidental global $U(1)$ symmetry in the
potential. Furthermore we study the phenomenology of the four Higgs
sector analytically and numerically and demonstrate that the VEV
configurations chosen in the numerical examples of the fermion mass
matrices cannot be minima of the potential, if CP is only
spontaneously violated. In the case of explicit CP-violation a
numerical analysis indicates the possibility that the chosen VEV
configurations can be minima of the general \Groupname{D}{5} invariant
potential. The \Groupname{D}{5} invariant three Higgs sector as well
as the four Higgs sector are compared with the corresponding
Higgs sectors invariant under the dihedral groups \Groupname{D}{3},
\Groupname{D}{4} and \Groupname{D}{6}. Thereby we
show the importance to classify the symmetries according to their
product structure rather than to pick one freely.

\noindent Finally, we briefly mention the possible embeddings of our
minimal model into GUT groups and continuous flavor groups.

\noindent The paper is organized as follows: Section 2 contains the
group theory of the dihedral symmetries. Our minimal model is presented
in Section 3 and the numerical analysis in Section 4. Section 5 is
dedicated to the Higgs sectors of \Groupname{D}{5} and the differences
to \Groupname{D}{3}, \Groupname{D}{4} and
\Groupname{D}{6}. Section 6 contains possible extensions of our model
from a low to a high energy theory. Finally, we conclude in Section 7
and comment on non-trivial subgroups of \Groupname{D}{5}. Clebsch Gordan coefficients and
embeddings of \Groupname{D}{5} are delegated to Appendix A. Appendix B
lists the numerical solutions for the Yukawa couplings and Higgs VEVs and
Appendix C contains the used experimental data.

\section{Group Structure of Dihedral Groups}
\label{sec:grouptheory}

\mathversion{bold}
\subsection{General Properties of Dihedral Groups \Groupname{D}{n}}
\mathversion{normal}
\label{sec:grouptheorydn}

The groups \Groupname{D}{n} are well-known in solid state and
molecular physics. Their double-valued counterparts are the groups
\Doub{D}{n}. Since $n \in \mathbb{N}$, there are infinitely many of
them. Apart from the two trivial groups with
$n=1,2$ all groups \Groupname{D}{n} are non-abelian. They only contain real one- and
two-dimensional irreducible representations. If its index $n$ is even, the group
\Groupname{D}{n} has four one- and $\frac{n}{2}-1$
two-dimensional representations and for $n$ being odd \Groupname{D}{n} has
two one- and $\frac{n-1}{2}$ two-dimensional representations. The
order of the group \Groupname{D}{n} is $2 \, n$. 
The four smallest non-abelian discrete groups can be found among
the family of the dihedral symmetries: \Groupname{D}{3},
\Groupname{D}{4}, \Doub{D}{2} and \Groupname{D}{5}. Generators of the
two-dimensional representations can be given for all $n$ \cite{Lomont}:
\begin{equation}\label{generators}
\rm A =\left(\begin{array}{cc} 
                           \mathrm{e}^{\left( \frac{2 \pi i}{n} \right) \, j} & 0 \\
                            0 & \mathrm{e}^{-\left( \frac{2 \pi i}{n} \right)
                              \, j} 
          \end{array}\right) \;\;\; , \;\;\; \rm B=\left(\begin{array}{cc} 
                                       0 & 1 \\
                                       1 & 0 
                  \end{array}\right) 
\end{equation}
with $j=1, \dots, \frac{n}{2} -1$ for $n$ even and $j=1, \dots,
\frac{n-1}{2}$ for $n$ odd. They fulfill the relations:
\begin{equation}\label{genrelations}
\mathrm{A}^{n} =\mathbb{1} \;\;\; , \;\;\; \rm B^2=\mathbb{1} \;\;\; ,
\;\;\; \rm ABA=B \; . 
\end{equation}
The corresponding character tables can also be found in
\cite{Lomont}. Note that we have chosen complex
generators for the two-dimensional representations. Since these are real,
there exists a unitary matrix $U$ which links their generators to its complex conjugates:
$U= \left( \begin{array}{cc} 0 & 1  \\ 1 & 0 \end{array} \right)$. 
For any $\left( \begin{array}{c} a_{1}
    \\ a_{2} \end{array} \right) \sim \Rep{2}$ the combination $ U \, \left( \begin{array}{c} a_{1} ^{\star}
  \\ a_{2} ^{\star} \end{array} \right) = \left( \begin{array}{c} a_{2} ^{\star}
  \\ a_{1} ^{\star} \end{array} \right)$ transforms as $\Rep{2}$
instead of $ \left( \begin{array}{c} a_{1} ^{\star}
  \\ a_{2} ^{\star} \end{array} \right)$, as it would be the case for real
generators $\rm A$ and $\rm B$.

\mathversion{bold}
\subsection{The Group \Groupname{D}{5}}
\mathversion{normal}
\label{sec:grouptheoryd5}
\Groupname{D}{5} is of order ten and has two one- and two two-dimensional
irreducible representations, since its index is odd. They are denoted
as $\MoreRep{1}{1}$, $\MoreRep{1}{2}$, $\MoreRep{2}{1}$ and
$\MoreRep{2}{2}$. Both two-dimensional representations are
faithful. Their characters $\chi$,
i.e. the traces of their representation matrices, are given in the
character table, shown in \Tabref{chartabD5}. There we use the following notations: $\Cl{i}$ with
$i=1,...,4$ are the four classes of the
group, $\OrdCl{i}$ is the order of the $i ^{\mathrm{th}}$ class,
i.e. the number of distinct elements contained in this class, $\Ord{h}{\Cl{i}}$
is the order of the elements $R$ in the class $\Cl{i}$, i.e. the smallest
integer ($>0$) for which the equation $R ^{\Ord{h}{\Cl{i}}}= \mathbb{1}$
holds. Furthermore the table contains one representative for each
class $\Cl{i}$ given as product of the generators $\rm
A$ and $\rm B$ of the group. The elements belonging to the classes
$\Cl{i}$ are: $\Cl{1} = \left\{ \mathbb{1} \right\}$, $\Cl{2} = \rm
\left\{ B, B A, B A^2, B A^3, B A^4 \right\}$, $\Cl{3} = \rm \left\{ A,
  A^4 \right\}$ and $\Cl{4} = \rm \left\{ A^2, A^3 \right\}$. With the
help of the character table the Kronecker
products can be calculated. They are $\MoreRep{1}{i} \times \MoreRep{1}{j} = \MoreRep{1}{\scriptsize \rm
  (i+j) mod 2 +1 \normalsize}$ , $\MoreRep{1}{i} \times \MoreRep{2}{j} =
\MoreRep{2}{\rm j}$ for $\left\{ \rm i, \rm j
\right\} \in \left\{ 1,2 \right\}$ and $\MoreRep{2}{1} \times
\MoreRep{2}{2} = \MoreRep{2}{1} + \MoreRep{2}{2}$ and $\left[
  \MoreRep{2}{i} \times \MoreRep{2}{i}\right] = \MoreRep{1}{1} +
\MoreRep{2}{j}$ , $\left\{ \MoreRep{2}{i} \times
  \MoreRep{2}{i}\right\} = \MoreRep{1}{2} \;\; \mbox{for} \; \rm i
\neq \rm j$  where  $\left[ \mu \times \mu \right]$ is the symmetric part of the product $\mu \times \mu$
and $\left\{ \mu \times \mu \right\}$ the anti-symmetric one. Note further that $\mu \times \nu =
\nu \times \mu$ for all representations $\mu$ and $\nu$. Taking $n=5$
in \Eqref{generators} and \Eqref{genrelations}
gives the generators $\mathrm A$ and $\mathrm B$ and their relations
for \Groupname{D}{5}. They are required for the calculation of
the Clebsch Gordan coefficients being shown in \Appref{app:grouptheory}. They
actually coincide with the matrices chosen in \cite{d5}. The embedding of \Groupname{D}{5} into continuous groups is
very interesting with respect to grand unified model
building. Therefore we show how \Groupname{D}{5} can be embedded into $SO(3)$ and
$SU(3)$ in \Appref{app:grouptheory}. We will discuss this in more
detail in \Secref{sec:extensions}.

\begin{table}
\begin{center}
\begin{tabular}{|l|cccc|}
\hline
&\multicolumn{4}{|c|}{classes}                                                 \\ \cline{2-5}
&$\Cl{1}$&$\Cl{2}$&$\Cl{3}$&$\Cl{4}$\\
\cline{1-5}
\rule[0.15in]{0cm}{0cm} $\rm G$                 &$\rm \mathbb{1}$&$\rm B$&$\rm A$               &$\rm A^{2}$                    \\
\cline{1-5}
$\OrdCl{i}$          &1      &5      &2                      &2                      \\
\cline{1-5}
$\Ord{h}{\Cl{i}}$                   &1      &2      &5                      &5            \\
\hline
\rule[0in]{0.3cm}{0cm}$\MoreRep{1}{1}$                        &1      &1      &1      &1\\                             
\rule[0in]{0.3cm}{0cm}$\MoreRep{1}{2}$                        &1      &-1     &1      &1\\                             
\rule[0in]{0.3cm}{0cm}$\MoreRep{2}{1}$                        &2      &0      &$\alpha$       &$\beta$\\
\rule[0in]{0.3cm}{0cm}$\MoreRep{2}{2}$                        &2      &0      &$\beta$        &$\alpha$\\[0.1cm]
\hline
\end{tabular}
\end{center}
\begin{center}
\begin{minipage}[t]{12cm}
\caption[]{Character table of the group
  \Groupname{D}{5}. $\alpha$ and $\beta$ are given as $\alpha=
  \frac{1}{2} \, \left(-1 +\sqrt{5} \right) = 2 \, \cos (\frac{2 \pi}{5})$
  and $\beta= \frac{1}{2} \, \left(-1-\sqrt{5} \right) = 2 \, \cos
  (\frac{4 \pi}{5})$ and therefore $\alpha+ \beta=-1$. For further
  explanations see text. \label{chartabD5}}
\end{minipage}
\end{center}
\end{table}

\section{Minimal Model}
\label{sec:model}
Here we present a minimal model which leads to viable mass spectra
and mixing parameters for quarks as well as leptons with the
Higgs potential being free from accidental symmetries (see
\Secref{sec:Higgspotentials}). We assign the left-handed quarks $Q
_{i}= \left( u _{i}, d_{i} \right) _{L} ^{T}$ and their conjugates
$u^{c} _{L \, i}$, $d^{c} _{L \, i}$ of the $i ^{\rm th}$
generation to: 
\[
Q_{1} \sim \MoreRep{1}{1} \; ; \;\; \left( \begin{array}{c} Q_{2} \\
 Q_{3}  \end{array} \right) \sim \MoreRep{2}{2} \; ; \;\; u^{c} _{L \,
1} \, ,
\, d^{c} _{L \, 1} \sim \MoreRep{1}{1} \;\;\; \mbox{and} \;\;\; \left(
  \begin{array}{c} u^{c} _{L \, 2} \\ u^{c} _{L \, 3} \end{array} \right) \, ,
\, \left(
  \begin{array}{c} d^{c} _{L \, 2} \\ d^{c} _{L \, 3} \end{array} \right)\sim
\MoreRep{2}{1} \; .
\]
The left-handed lepton doublets $L_{i} = \left( \nu _{i} , e_{i}
\right) _{L} ^{T}$ and its conjugates
$e^{c} _{L \, i}$ and $\nu^{c} _{L \, i}$ transform in a similar way, i.e.:
\[
L_{1} \sim \MoreRep{1}{1} \; ; \;\; \left( \begin{array}{c} L_{2} \\
 L_{3}  \end{array} \right) \sim \MoreRep{2}{2} \; ; \;\; e^{c} _{L \,
1} \, ,
\, \nu^{c} _{L \, 1} \sim \MoreRep{1}{1} \;\;\; \mbox{and} \;\;\; \left(
  \begin{array}{c} e^{c} _{L \, 2} \\ e^{c} _{L \, 3} \end{array} \right) \, ,
\, \left(
  \begin{array}{c} \nu^{c} _{L \, 2} \\ \nu^{c} _{L \, 3} \end{array}
\right)\sim \MoreRep{2}{1} \; .
\]
The four Higgs fields $\chi _{i}$ and $\psi _{i}$ which are $SU(2)
_{L}$ doublets with hypercharge $Y=-1$ (like the Higgs field in the
SM) transform as $\left( \begin{array}{c} \chi _{1} \\ \chi _{2} \end{array} \right) \sim
\MoreRep{2}{1}$ and $\left( \begin{array}{c} \psi _{1} \\ \psi _{2} \end{array} \right) \sim
\MoreRep{2}{2}$ under \Groupname{D}{5}.\\

\noindent The fermion mass
  matrices arise from the coupling $y_{ij} \, \mathrm{L}^{T} _{i} \,
  \epsilon \, \xi \, \mathrm{L}^{\mathrm{c}} _{j}$ for down-type
  quarks ($\mathrm{L}_{i}=Q_{i}$, $\mathrm{L} ^{\mathrm{c}}
    _{i}=d^{c} _{L \, i}$) and charged leptons ($\mathrm{L}_{i}= L_{i}$,
    $\mathrm{L} ^{\mathrm{c}} _{i}= e^{c} _{L \, i}$) and $y_{ij} \, \mathrm{L}^{T} _{i} \,
  \epsilon \, \tilde{\xi} \, \mathrm{L}^{\mathrm{c}} _{j}$ for up-type
  quarks ($\mathrm{L} _{i}= Q_{i}$, $\mathrm{L}
  ^{\mathrm{c}}_{i}=u^{c} _{L \, i}$) and neutrinos ($\mathrm{L} _{i}=
  L_{i}$, $\mathrm{L} ^{\mathrm{c}} _{i} = \nu ^{c} _{L \, i}$). Thereby, the Higgs field $\xi$ is $\xi ^{T}= (
  \xi^{0} \; , \;
 \xi ^{-} ) ^{T}$ and its complex conjugate $\tilde{\xi}$ is
 $\tilde{\xi} = \epsilon \, \xi^{\star}$ with $\epsilon$ being the
 anti-symmetric 2-by-2 matrix in $SU(2)_L$ space and the star
 $^{\star}$ denotes the complex conjugation.

\noindent The resulting Dirac mass matrices are:\\
\small
\rule[0in]{-0.5in}{0in}\parbox{7in}{\begin{equation}\label{Diracmasses}
\mathcal{M} _{u, \nu} = \left( \begin{array}{ccc} 
    0 &  \alpha ^{u,\nu} _{2} \, \langle \chi _{1} \rangle ^{\star} &
    \alpha ^{u , \nu} _{2} \, \langle \chi _{2} \rangle ^{\star} \\
    \alpha ^{u , \nu} _{3} \, \langle \psi _{1} \rangle ^{\star} &
    \alpha ^{u , \nu} _{1} \, \langle \psi _{2} \rangle ^{\star} &
    \alpha ^{u , \nu} _{0} \, \langle \chi _{1} \rangle ^{\star} \\
    \alpha ^{u , \nu} _{3} \, \langle \psi _{2} \rangle  ^{\star} & 
    \alpha ^{u , \nu} _{0} \, \langle \chi _{2} \rangle ^{\star} &
    \alpha ^{u , \nu} _{1} \, \langle \psi _{1} \rangle ^{\star}
    \end{array}
    \right) , \, \mathcal{M} _{d, l} = \left( \begin{array}{ccc} 
    0 & \alpha ^{d,l} _{2} \, \langle \chi _{2} \rangle & \alpha
    ^{d,l} _{2} \, \langle \chi _{1} \rangle  \\
    \alpha ^{d,l} _{3} \, \langle \psi _{2} \rangle & \alpha ^{d,l}
    _{1}  \, \langle \psi _{1} \rangle &  \alpha
    ^{d,l} _{0} \, \langle \chi _{2} \rangle  \\
    \alpha ^{d,l} _{3} \, \langle \psi _{1} \rangle &  \alpha ^{d,l}
    _{0} \, \langle \chi _{1} \rangle & \alpha ^{d,l} _{1} \, \langle \psi
    _{2} \rangle 
    \end{array}
    \right) \; ,
\end{equation}}\\
\normalsize
where $\langle \xi \rangle$ denotes the VEV of the field $\xi=
\psi_{i}, \chi_{i}$.  The
VEVs and the Yukawa couplings $\alpha ^{u,d,l,\nu} _{j}$ are in
general complex. The (1,1) element of the mass matrices is zero, since there
is no Higgs field transforming trivially under \Groupname{D}{5}. Even though there are more parameters in our model than
observables to fit, this is a rather 
non-trivial task, since apart from the number of free parameters also
the structure of the mass
matrices plays an important role in fitting the
observables.

\noindent The number of parameters could obviously be reduced, if some
of the Yukawa couplings were assumed to be equal. Since our flavor
symmetry \Groupname{D}{5} cannot explain this, we do
not use such assumptions. Another way to reduce the number of parameters
could be to set some of the VEVs to be equal or zero. For two VEVs being
zero we either have two massless quarks or cannot generate
CP-violation, since $\mathcal{J} _{CP} \propto
\mbox{det} \left( \left[ \mathcal{M}_{u} \, \mathcal{M}_{u} ^{\dagger} ,
  \mathcal{M}_{d} \, \mathcal{M}_{d} ^{\dagger} \right] \right)$
\cite{JCPrel} vanishes. Furthermore some of these configurations lead to the
appearance of accidental symmetries in the Higgs potential (see
\Secref{sec:Higgs4}). For one VEV being zero or two VEVs being equal
we cannot find an obvious reason to exclude these assumptions, but one
does not gain much in doing so, since most of the free
parameters in our model come from the (in total) 16 Yukawa couplings which have to
be compared with the 20 (22) observable masses and mixing parameters
in the quark and lepton sector for Dirac (Majorana)
neutrinos. Therefore we do not make such assumptions in the following
numerical study.

\noindent We have chosen a structure which is similar to a mass
texture which has already been discussed in
the literature \cite{masstexture}. It actually arises from our mass
matrix for real parameters and in the limit that all VEVs are equal in
\Eqref{Diracmasses} together with $\alpha _{2} ^{i} = \alpha _{3}
^{i}$ for $i=u,d,l,\nu$ or for $\langle \chi_{1} \rangle = \langle
\chi_{2} \rangle$, $\langle \psi_{1} \rangle = \langle \psi_{2}
\rangle$ and $\alpha _{3} ^{u , \nu} = \frac{\langle \chi_{2} \rangle ^{\star}}{\langle
 \psi_{2} \rangle ^{\star}} \, \alpha _{2} ^{u , \nu}$, $\alpha _{3}
^{d , l} = \frac{\langle \chi_{2} \rangle}{\langle \psi_{2} \rangle} \,
\alpha _{2} ^{d , l}$ . Then all mass matrices are invariant
under the interchange of the second and third generation which always
leads to mixing angles
$\theta_{13}=0$ and $\theta_{23}= \frac{\pi}{4}$ with unconstrained
$\theta_{12}$. In the leptonic sector this is called
$\mu - \tau$ interchange symmetry \cite{mutau}. 

\noindent Here we assume that the first generation transforms as
$\MoreRep{1}{1}$ and the second and third one as $\MoreRep{2}{i}$ under
\Groupname{D}{5}. This choice is inspired by the observation that the
masses of the particles which belong to the first generation are much
smaller than the masses of the ones of the second and third one and by
the fact that the mixing in the 2-3 sector of the leptons is large,
possibly maximal. In general there are six possibilities to assign the
three generations $\left\{ 1,2,3 \right\}$ to $\MoreRep{1}{i} + \MoreRep{2}{j}$: $\left\{
  \left[ 1 \right], \left[ 2 , 3 \right] \right\}$, $\left\{
  \left[ 1 \right], \left[ 3 , 2 \right] \right\}$, $\left\{
  \left[ 2 \right], \left[ 1 , 3 \right] \right\}$, $\left\{
  \left[ 2 \right], \left[ 3 , 1 \right] \right\}$, $\left\{
  \left[ 3 \right], \left[ 1 , 2 \right] \right\}$ and $\left\{
  \left[ 3 \right], \left[ 2 , 1 \right] \right\}$ where $\left[ .
\right]$ forms the one-dimensional representation and $\left[ . ,
  . \right]$ the two-dimensional one under \Groupname{D}{5}. If the
left-handed fields are permuted by $P$: $P \, \left\{ \left[1 \right],
\left[ 2, 3 \right] \right\} = \left\{ \left[ P_{1} \cdot \left( 1,2,3 \right)
^{T} \right],  \left[ P_{2} \cdot \left( 1,2,3 \right) ^{T},  P_{3}
\cdot \left( 1,2,3 \right) ^{T} \right] \right\}$ with $P_{i}$
being the $i ^{\mathrm{th}}$ row of the matrix $P$ and the left-handed
conjugate fields by $Q$, the mass matrix $\mathcal{M}$ changes to $\mathcal{\tilde{M}}=P
\, \mathcal{M} \, Q^{T}$, since all permutations are orthogonal. As
one can see these permutations do neither change the eigenvalues of
the mass matrix, i.e. $\mathrm{det} (\mathcal{M} \, \mathcal{M}
^{\dagger} - \Lambda \, \diag (1,1,1))=0$ remains invariant, nor the
mixing matrices $V_{CKM}$ and $U_{MNS}$. If the mass matrix of the
up-type quarks $\mathcal{M}_{u}$ is diagonalized by $U_{u}$ fulfilling
$U_{u}^{\dagger} \, \mathcal{M}_{u} \, \mathcal{M} _{u} ^{\dagger} \,
U_{u} = \diag (m_{u} ^{2}, m_{c} ^{2}, m_{t} ^{2})$ and the same holds
for $\mathcal{\tilde{M}}_{u}$ and $\tilde{U} _{u}$, then $U_{u}$ and
$\tilde{U} _{u}$ are connected by $\tilde{U} _{u}=P \,
U_{u}$. Similarly one gets for
the down-type quarks $\tilde{U} _{d}= P \, U_{d}$ \footnote{The permutations
have to be both $P$, since $u_{L}$ and $d_{L}$ transform in the same
representation of the SM.} and therefore, for example, $V_{CKM}= \tilde{U} _{u} ^{T} \,
\tilde{U} _{d} ^{\star} = U_{u} ^{T} \, P^{T} \, P^{\star} \, U_{d}
^{\star}= U_{u} ^{T} \, U_{d} ^{\star}$, such that $V_{CKM}$ is not
affected by this permutation. For the mass matrix texture it seems to
be most convenient to have a vanishing $(1,1)$ element instead
of, for example, a $(2,3)$ or $(3,3)$ one.

\noindent Apart from permuting the three generations among each other
one can interchange the transformation properties of the left-handed
and left-handed conjugate fields. This leads to matrices which are
transposed to the ones shown in \Eqref{Diracmasses}. Furthermore one
can ask whether there is a considerable change, if the first
generation is not assigned to $\MoreRep{1}{1}$, but to
$\MoreRep{1}{2}$. The answer is no, since it only introduces a
relative sign between the $(1,2)$ and $(1,3)$ and $(2,1)$ and $(3,1)$
elements of the mass matrix.

\noindent Our choice for the assignment of fermion
generations allows an embedding into the Pati-Salam gauge group,
where all left-handed fields are unified into one
representation as well as all left-handed conjugate fields into the
conjugated one. One can also attempt to embed the model into $SO(10)$, but
then all fermions have to transform in the same way under
\Groupname{D}{5}. In doing so one arrives at mass matrices which have
two additional texture zeros in the (2,3) and (3,2) element (and one
Yukawa coupling less than the matrices shown above). In
case of hermitian matrices such a texture is excluded for
quarks \cite{4texturezeros}. This does not strictly apply in our
case, because our matrices are in general not hermitian, but we
believe that this does not change the result of
\cite{4texturezeros}. For an embedding into $SU(5)$, the generations $Q _{i}$ and
$u^{c} _{L \, i}$ would have to transform in the same way under
\Groupname{D}{5}, since these fields are unified into the 10-plet of
$SU(5)$. Then again the mass matrix for the up-type quarks has to have
three texture zeros in the positions (1,1), (2,3) and (3,2) combined
with a mass matrix for the down-type quarks with one zero in the (1,1)
element, since $Q_{i}$ and $d_{L \, i} ^{c}$ do
not belong to the same $SU(5)$ representation and therefore can
transform differently under \Groupname{D}{5}. Generally, such a structure is not excluded, but taking into
account the various relations among the non-vanishing matrix elements,
it seems to be unfavorable. Therefore we do not discuss this
possibility here. Concerning the number of possible different
assignments for quarks and leptons the Pati-Salam group has an
advantage over $SU(5)$, since the sixteen fermions of one generation
(i.e. the right-handed neutrino is always included in our
considerations) are unified into two and not into three
representations of the gauge group. Its disadvantage is the fact that
the three SM gauge factors are not unified into a single group, but
rather in a product one.

\noindent If neutrinos are Majorana particles, the Majorana mass
matrix for the right-handed neutrinos looks very simple, since our model does not include SM
gauge singlets transforming non-trivially under \Groupname{D}{5}:
\small
\begin{equation}\label{Majoranamass}
\mathcal{M} _{RR} = \left( \begin{array}{ccc} 
    M_{1} & 0 & 0\\
    0 &  0 & M_{2}\\
    0 &  M_{2} & 0
    \end{array}
    \right) \; .
\end{equation}
\normalsize
The resulting mass matrix for the light neutrinos is then given
through the type I seesaw \cite{seesaw1} formula
\begin{equation}
M_{\nu} = (-) \mathcal{M}_{\nu} \, \mathcal{M} _{RR} ^{-1} \,
\mathcal{M}_{\nu} ^{T} \; .
\end{equation}
As one can see, two of the right-handed neutrinos are degenerate at
tree-level. This can be used for resonant leptogenesis \cite{reslepto}. 

\noindent An important aspect of our symmetry driven discussion is
that different from the usual
assumption in papers treating a certain texture of the mass matrices
(like \cite{masstexture}) the Majorana mass matrix for the
right-handed neutrinos strongly differs from the structure of the
Dirac masses. Therefore also the effective mass matrix for the
light neutrinos is in general distinct from the (Dirac) mass matrices of
the other fermions. The reason for this simply lies in the fact that
Majorana and Dirac masses do arise from completely different
mechanisms with different symmetry aspects: first the Dirac masses connect different fields whereas
Majorana masses connect the same field with itself and second Dirac
masses arise through the coupling of $SU(2) _{L}$ doublet Higgs fields
with hypercharge $Y = \pm 1$ unlike Majorana masses which are direct
mass terms for right-handed neutrinos and are mediated by $SU(2) _{L}$
Higgs triplets for left-handed ones.

\noindent \Groupname{D}{5} has two distinct two-dimensional
representations instead of only one like \Groupname{D}{3}. The
differences in the mass matrices which follow from this fact will be
studied next. In \cite{d3texture} the authors assigned the
fermion generations and three Higgs fields to $\MoreRep{1}{1} +
\Rep{2}$ under \Groupname{D}{3}. We observe that we cannot use the
  same representation structure in the Higgs sector in our
  \Groupname{D}{5} model for a realistic theory due to an accidental
  $U(1)$ symmetry in the potential (see \Secref{sec:Higgs3}). If we do so anyway, we can
  distinguish two cases in \Groupname{D}{5}: both fermion generations
  and Higgs fields transform as $\MoreRep{1}{1} + \MoreRep{2}{i}$
  under \Groupname{D}{5} or the fermions are in $\MoreRep{1}{1} +
  \MoreRep{2}{i}$ and the Higgs fields are in $\MoreRep{1}{1} +
  \MoreRep{2}{j}$ with $\rm i \neq j$. In the first case the
  \Groupname{D}{5} invariance leads to a mass matrix with two zeros on
  its diagonal, i.e. the $(2,2)$ and $(3,3)$ element vanish, since
  $\MoreRep{2}{i} \times \MoreRep{2}{i}$ does not contain
  $\MoreRep{2}{i}$ for $\rm i=1,2$ in contrast to $\Rep{2} \times \Rep{2}
  \ni \Rep{2}$ in \Groupname{D}{3}. In the latter case the first
  generation transforming trivially under \Groupname{D}{5} is
  decoupled from the two other ones forming a two-dimensional
  representation, since $\MoreRep{1}{1} \times \MoreRep{2}{i} =
  \MoreRep{2}{i}$ for $\rm i=1,2$.

\noindent Thus the existence of two two-dimensional representations in the
flavor group has two main consequences on the structure of the mass
matrices: on the one hand it tends to reduce the number of allowed
Yukawa couplings and so maintaining texture zeros becomes easier, on
the other hand it leaves the freedom of assigning the three
generations of fermions to different two-dimensional representations
(as done here).

\section{Phenomenological Analysis}
\label{sec:numerics}

One appropriate example for a starting point of our numerical analysis
is given by
\begin{equation}
\label{startpoint}
\mathcal{M} _{start} = \left( \begin{array}{ccc}
    0 & 0 & 0\\
    0 & a & b\\
    0 & b & a
    \end{array}
    \right) \; .
\end{equation}
With this matrix one can already fit the masses of the second
and third generation fermions by fixing $a$ and $b$. The eigenvalues
of $\mathcal{M}_{start}$ are $\left( 0, a-b, a+b \right)$. The mass of the
third generation can be taken to be $a-b$ and the one of the second
one $a+b$. It is clear then that $\mathrm{sign}(a)= -
\mathrm{sign}(b)$. The mass of the third generation determines the absolute values
of $a$ and $b$ and the second generation the difference of $|a|$ and
$|b|$. The vanishing eigenvalue of $\mathcal{M}_{start}$ also explains the
smallness of the first generation compared to the two other ones. Such
a matrix is closely connected to the mass matrix of the light
neutrinos for $b \neq 0$ \cite{23blockmass,mutau} where it leads to
maximal atmospheric mixing. Although it contains this large mixing angle we
can use it for the description of quarks, because taking this form
for up-type as well as down-type quark mass matrices makes the two large mixing angles
cancel such that the angle $\theta_{23}$ can be arbitrarily small in this sector.

\noindent The matrix in \Eqref{startpoint} arises from \Eqref{Diracmasses} for $\langle
\chi_{1} \rangle = \langle \chi_{2} \rangle$, $\langle \psi_{1} \rangle
= \langle \psi_{2} \rangle$ and $\alpha _{2,3}^{i} =0$ for $i=u,d,l,\nu$. As argued in \Secref{sec:Higgs4}
one can arrange the Higgs potential to have an extremum for VEVs being
pairwise equal. Since the difference of $|a|$
and $|b|$ is determined by the mass of the second generation, $|a|
\approx |b|$ holds. This can be maintained if all VEVs are nearly
equal and $|\alpha _{0} ^{i}| \approx |\alpha _{1} ^{i}|$. As
shown in \Secref{sec:Higgs4} also this is allowed by the
minimization conditions. Note that \Groupname{D}{5} does
not restrict the Yukawa couplings $\alpha _{i} ^{j}$. Therefore $|\alpha
_{0} ^{i}| \approx |\alpha _{1} ^{i}|$ is not favored by the flavor
symmetry. Also our assumption $|\alpha _{2,3} ^{i}| \ll |\alpha _{0,1}
^{i}|$ is not guaranteed by any symmetry of the model. In order to
achieve this, one could for
example introduce a $U(1)_{FN}$ factor acting non-trivially in flavor
space to implement the Froggatt Nielsen (FN) mechanism
\cite{fnmechanism}. We could assign a non-vanishing charge $+q$ to the
first generation and let the second and third generation be neutral
under this $U(1)_{FN}$. We then gain a suppression factor of $\epsilon
^{q}$ with $\epsilon \equiv \frac{\langle \theta \rangle}{M}$ for the
matrix elements of the first row and column compared to the
others. $\langle \theta \rangle$ is the VEV of the scalar SM gauge
singlet $\theta$ having charge $-1$ under $U(1)_{FN}$ and $M$ is the mass
of some vector-like fermions. These fields are assumed to be very
heavy and therefore actually decouple from our low energy theory. Note
that the second and third generation of fermions have to transform in
the same way under the $U(1)_{FN}$, since otherwise the $U(1)_{FN}$ would not
commute with our flavor symmetry \Groupname{D}{5}. Note further
that the zero in the $(1,1)$ element is independent of the FN
mechanism, since it comes from our assignment of fermions and Higgs
fields under \Groupname{D}{5}.\\

\noindent Next we present our numerical examples for Dirac and
Majorana neutrinos. As already stated above, the mass matrices contain
in general too many parameters to make predictions. In order to reduce
the number of free parameters we restrict ourselves to real Yukawa couplings and allow
the VEVs $\langle \chi_{1,2} \rangle$ and $\langle \psi_{1} \rangle$
to have non-vanishing (complex) phases. We show the numerical values of the
Yukawa couplings and VEVs in \Appref{app:tables}. With these the best
fit values of the measured quantities shown in \Appref{app:experimentaldata} can
be accommodated within the given error bars. Interestingly, all phases
of the VEVs turn out to be
small. Although the Yukawa couplings are chosen to be real, SCPV is
excluded, since the parameters in the Higgs
sector have to be complex in order to allow the shown VEV
configurations to be minima of the potential. This fact will be
explained in detail in \Secref{sec:Higgs4}. The mass ordering of the (light)
neutrinos is normal in both examples. This is not a general feature of
our model, but rather chosen by us for simplicity. As we can fit all measured
quantities, we only discuss the results for the unmeasured ones.\\
\noindent In case of Dirac neutrinos the sum of the neutrino masses is $0.2255
\eV$. This is below the current bound obtained from cosmology, even if
the Lyman $\alpha$ data are included \cite{cosmomassbounds}. However, it will be
measurable in the next five to ten years \cite{cosmomassprospect}. $s_{13} ^{2}$ is
about $0.012$ and hence a factor of three below the current CHOOZ
bound, but detectable quite soon in the next generation of
reactor experiments \cite{theta13prospect}. The Dirac
phase $\delta$ is $\sim 3.6$ radian. The quantity $m_{\beta}$ measured
in beta decay experiments is $0.07 \eV$. This is below the current
limit of $2.2 \eV$ \cite{betadecaybounds} and also a factor of three below the one of
the planned KATRIN experiment \cite{betadecayprospect}.\\
\noindent Before presenting the corresponding results in case of
Majorana neutrinos, we comment on the generic problem of Dirac
neutrinos. As one can see in \Appref{app:tables} the Yukawa couplings of
the neutrinos $\alpha ^{\nu} _{i}$ have to be suppressed by nine to
twelve orders of magnitude compared
to the other fermions to ensure that the neutrinos have masses of the
order $1 \eV$. Clearly, our flavor symmetry \Groupname{D}{5} does not
explain this, but an additional $U(1) _{FN}$ family symmetry can do
so. If the right-handed neutrinos have a charge $\sim q_{f} + 10$
under $U(1) _{FN}$ where $q_{f}$ is the charge of any other fermion
under $U(1)_{FN}$, the neutrino couplings can be suppressed by an
additional factor $\epsilon ^{10}$. For $\epsilon \sim 0.1$ this gives
the right order of magnitude for the neutrino masses. However, then
the model cannot be embedded into the Pati-Salam group.\\
\noindent Next we consider the neutrinos to be Majorana particles. In
this case the type I seesaw \cite{seesaw1} explains the smallness of
the neutrino masses without an extra suppression of their Yukawa
couplings $\alpha ^{\nu} _{i}$. The masses for the light neutrinos are
$(0.1146,0.1149,0.1242) \eV$. The sum of their masses is therefore
still below the current bound, but could be measured by the planned
experiments. The scale of the right-handed neutrino masses is about
$10^{14} \GeV$, but it can be rescaled by proper redefinition of the
neutrino Yukawa couplings $\alpha ^{\nu} _{i}$. Interestingly, $s_{13}
^{2}$ is around the $2 \, \sigma$ limit of the CHOOZ experiment. The
CP phases which are not constrained by experiments are $(\delta,
\varphi_{1}, \varphi_{2}) \sim (3.9,0.74,0.33)$ radian. $m_{\beta}$ is
- similar to the Dirac case - a factor of two smaller than the bound
which can be obtained by the
KATRIN experiment. $|m_{ee}|$ which is measured in neutrinoless double
beta decay is about $0.1 \eV$. This is an order of magnitude below the
upper bound \cite{0vbbbounds}, but can be measured in the next five to ten
years \cite{0vbbprospect}.\\
\noindent The smallness of $m_{\beta}$ and $|m_{ee}|$ is due to the
normal ordering of the (light) neutrinos. Finally, we summarize all
mentioned quantities in \Tabref{results}.
\begin{table}[h!]
\begin{center}
\small
\begin{tabular}{|c|c|c|}
\hline
Quantity & Dirac neutrinos & Majorana neutrinos \\
\hline
$m_{1}$ [$\eV$]& $0.0701$ & $0.1146$\\
$m_{2}$ [$\eV$]& $0.0706$ & $0.1149$\\
$m_{3}$ [$\eV$]& $0.0848$ & $0.1242$\\
$\sum _{i} \, m_{i}$ [$\eV$]& $0.2255$ & $0.3537$\\
\hline
$M_{R \, 1}$ [$\GeV$]& - & $1.878 \times 10^{14}$\\
$M_{R \, 2,3}$ [$\GeV$]& - & $2.011 \times 10^{14}$\\
\hline
$s_{13} ^{2}$& $0.0119$ & $0.0303$\\
\hline
$\delta$ [rad.]& $3.5775$ & $3.8619$\\
$\varphi_{1}$ [rad.]& - & $0.7396$\\
$\varphi_{2}$ [rad.]& - & $0.3312$\\
\hline
$m_{\beta}$ [$\eV$]& $0.0704$ & $0.1150$\\
\hline
$|m_{ee}|$ [$\eV$]& - & $0.1002$\\
\hline
\end{tabular}
\end{center}
\begin{center}
\begin{minipage}[t]{12cm}
\caption[]{ Numerical values for the unmeasured quantities of the
  leptonic sector. The Majorana phases
  $\varphi_{1,2}$ are given by the convention: $U_{MNS} = \tilde{V}
  _{CKM} \cdot \diag(\mathrm{e} ^{i \, \varphi_{1}}, \mathrm{e} ^{i \,
  \varphi_{2}},1)$ with $0 \leq \varphi_{1,2} \leq \pi$. \label{results}}
\end{minipage}
\end{center}
\end{table}
\normalsize

\mathversion{bold}
\section{Minimal Higgs Potentials in \Groupname{D}{5}}
\mathversion{normal}
\label{sec:Higgspotentials}

\subsection{Three Higgs Potential}
\label{sec:Higgs3}
In this Subsection we discuss the potential arising from the three Higgs
fields $\phi$, $\psi_{1}$ and $\psi_{2}$ where $\phi$ transforms as
any one-dimensional representation and $\left(
  \begin{array}{c} \psi_{1} \\ \psi_{2} \end{array} \right)$ forms
any doublet under \Groupname{D}{5}. The potential reads:
\small
\begin{eqnarray}
V _{3} (\phi, \psi_{i}) &=& -\mu_{1} ^{2} \, \phi
^{\dagger } \, \phi -\mu _{2} ^{2} \, \sum \limits _{i=1} ^{2}
\psi_{i} ^{\dagger}  \, \psi _{i} + \lambda _{s} \, \left( \phi
  ^{\dagger} \, \phi \right)^2 + \lambda _{1} \, \left( \sum \limits
  _{i=1} ^{2} \psi _{i} ^{\dagger} \, \psi_{i} \right) ^{2} \\
\nonumber
 &+&  \lambda _{2} \, \left( \psi _{1} ^{\dagger} \, \psi _{1}-\psi
   ^{\dagger} _{2} \, \psi _{2} \right) ^{2} + \lambda _{3} \, |
   \psi _{1} ^{\dagger} \, \psi _{2} |^{2} \\ \nonumber
 &+& \sigma _{1} \, \left( \phi ^{\dagger} \, \phi \right) \,
 \left(\sum \limits _{i=1} ^{2} \psi ^{\dagger} _{i} \, \psi _{i}
 \right) + \left\{ \sigma _{2} \, \left( \phi ^{\dagger} \, \psi
     _{1}\right) \, \left( \phi ^{\dagger} \, \psi _{2} \right) +
   \mathrm{h.c.} \right\} + \sigma _{3} \, \sum \limits _{i=1} ^{2} |\phi ^{\dagger} \, \psi _{i}|^{2}  
\end{eqnarray}
\normalsize
where only $\sigma_{2}$ is complex. It can be made real by appropriate
redefinition of the field $\phi$, for example. We want to show
that there exists an accidental $U(1)$ symmetry in this potential
apart from the gauge symmetry $U(1) _{Y}$. In order to see this let the Higgs fields
$\phi$ and $\psi _{i}$ transform as
\begin{equation}
\phi \;\;\; \rightarrow \;\;\; \mathrm{e} ^{i \, \alpha} \, \phi
\; , \;\;\;\; \psi _{1} \;\;\; \rightarrow \;\;\; \mathrm{e} ^{i \, \beta} \;
\psi _{1} \; , \;\;\;\; \psi _{2} \;\;\; \rightarrow \;\;\; \mathrm{e}
^{i \, \gamma} \, \psi _{2} \; .
\end{equation}
The only non-trivial condition for the phases $\alpha$, $\beta$ and
$\gamma$ arises from the term $\sigma_{2}$:
\begin{equation} \label{condsigma2}
2 \, \alpha-\beta-\gamma =0 \; ,
\end{equation}
i.e. $\alpha$ can be expressed as $\frac{1}{2} \, (\beta + \gamma)$
while $\beta$ and $\gamma$ can have any value. Consequently, there exist two $U(1)$
symmetries, called $U(1) _{\beta}$ and $U(1) _{\gamma}$, under which
the three fields have the charges: $Q(\phi ; \beta) = Q (\phi ;
\gamma) = \frac{1}{2}$, $Q (\psi _{1} ; \beta) = 1$, $Q (\psi _{1} ;
\gamma) = 0$ and vice versa for $\psi _{2}$: $Q(\psi _{2} ; \beta)
=0$, $Q (\psi _{2} ; \gamma) = 1$. Taking the two linear
independent combinations of the charges $Q (\chi ; Y) = - \left[ Q ( \chi ;
\beta) + Q (\chi ; \gamma) \right]$ and $Q (\chi ; X) = Q (\chi ; \beta)-Q
(\chi ; \gamma)$ for $\chi = \phi , \psi _{1} , \psi _{2}$ one
recovers the $U(1) _{Y}$ and a further $U(1) _{X}$ under which the two
fields $\psi _{i}$ transform with opposite charges and $\phi$ remains
invariant. Alternatively, the $U(1) _{X}$ could be defined such that
$Q  (\phi ; X) =-Q (\psi _{i} ; X)$ and $Q (\psi _{j} ; X) = 0$ with $i
\neq j$. Taking the first definition of the $U (1) _{X}$ charges
one sees that any non-vanishing VEV for a field $\psi _{i}$ leads to the
spontaneous breaking of the $U(1) _{X}$ and therefore to the
appearance of a massless Goldstone boson which is phenomenologically
unacceptable. There are two ways to circumvent this: first introduce
terms in the potential which explicitly break $U(1) _{X}$, but
also the \Groupname{D}{5} symmetry \footnote{This is similar to the soft breaking
terms invoked in the MSSM.} or second leave $U(1) _{X}$
unbroken. The first possibility increases the number of parameters by
at least four and is not explained in
terms of any (further) symmetry while the second one cannot be realized, if our
model should accommodate the fermion masses at tree-level without
further fields. Hence we abandon this three Higgs potential which
actually contains the minimal set of Higgs fields needed for the
construction of viable mass matrices. 

\noindent The accidental $U(1)$ symmetry found here becomes obvious in the
basis where the generators $\rm A$ and $\rm B$ of \Groupname{D}{5} are taken to be
the ones shown in \Eqref{generators}. If one chooses for example real
representation matrices (found in \cite{Lomont}), the resulting potential
still contains the extra $U(1)$, but it is rather non-trivial to show this.

\noindent If one sets $\sigma_2 =0$, the symmetry of the potential is
further increased to $U(1) ^{3}$, since then the condition
\Eqref{condsigma2} is no longer valid. The $U(1) ^{2}$ which then
exists in the $\left(\psi_{1}, \psi_{2} \right)$ space can be
enhanced to an $SU(2)$ by setting $\lambda_{3} = 4 \,
\lambda_{2}$. Then the terms $\lambda_{2}$ and $\lambda_{3}$ can be
written as $\sum _{a} \left( \Psi ^{\dagger} \tau_{a} \Psi \right)^2$
with $\Psi= \left( \psi_{1}, \psi_{2} \right) ^{T}$ and $\tau_{a}$ are
the Pauli matrices, i.e. it equals the invariant arising from $\left( \Rep{2}
\times \Rep{2} \right) ^{2} \ni \Rep{3} \times \Rep{3} \ni \Rep{1}$ in
$SU(2)$. $\sigma_{2}=0$ and/or $\lambda_{3}= 4 \, \lambda_{2}$ can be
enforced by the VEV conditions. One example for this is given by the
configuration where the VEVs of $\phi$ and $\psi_{1}$ are unequal zero
and $\langle \psi_{2} \rangle = 0$.

\noindent Let us comment on the origin of this accidental $U(1)$. For
this we compare our \Groupname{D}{5} invariant Higgs
potential to one being invariant under \Groupname{D}{3} and
\Groupname{D}{4}, respectively. The \Groupname{D}{3} invariant version
of our potential has already been discussed in the literature
\cite{d3pot}. Apart from the terms contained in the \Groupname{D}{5}
invariant potential it allows a further term, namely: $ \left\{ \tau
  \left[ \left( \phi ^{\dagger} \, \psi _{1} \right) \, \left( \psi
      _{2} ^{\dagger} \, \psi _{1} \right) \pm  \left( \phi ^{\dagger}
      \, \psi _{2} \right) \, \left( \psi _{1} ^{\dagger} \, \psi _{2}
    \right) \right] + \mathrm{h.c.} \right\}$ with $+$ for $\phi \sim
\MoreRep{1}{1}$ and $-$ for $\phi \sim \MoreRep{1}{2}$ (under \Groupname{D}{3}). This term is
\Groupname{D}{3} invariant, since the product $\Rep{2} \times
\Rep{2}$ contains the representation $\Rep{2}$ itself and therefore
$\Rep{2} \times \Rep{2} \times \Rep{2} \times \MoreRep{1}{i} \ni
\MoreRep{1}{1}$ for $\rm i=1,2$. In \Groupname{D}{5} the corresponding
coupling is of the form $\MoreRep{2}{i} \times \MoreRep{2}{i} \not \ni
\MoreRep{2}{i}$ for both $\rm i=1,2$. Clearly, the $\tau$ term does not
allow for a further $U(1)$ symmetry, since it enforces the relations
$2 \, \beta - \alpha - \gamma =0$ and $2 \, \gamma - \alpha - \beta=0$
for the phases $\alpha$, $\beta$ and $\gamma$. This term has to
vanish, if the potential should be invariant under the reflection
symmetry $\phi \; \rightarrow  \; - \phi$ and $\psi_{1,2} \;
\rightarrow \; \psi_{1,2}$ as mentioned in \cite{s3potreflect}. Then
there exists an accidental $U(1)$ which was already realized in \cite{S3reflectGB}.

\noindent To compare our potential to the one
being invariant under \Groupname{D}{4} one has to notice that the
product $\Rep{2} \times \Rep{2}$ decomposes into $\sum \limits _{\mathrm{i}=1} ^{4} \,
\MoreRep{1}{i}$ there. Hence the quartic coupling $\lambda_{3}$ has to
be replaced by 
\begin{equation}\label{Higgs3D4}
\lambda_{3} \, \left( \psi^{\dagger} _{1} \, \psi _{2}-\psi
  ^{\dagger} _{2} \, \psi _{1} \right) ^{2} + \tilde{\lambda} _{3} \, \left( \psi^{\dagger} _{1} \, \psi _{2} + \psi
  ^{\dagger} _{2} \, \psi _{1} \right) ^{2} \; .
\end{equation}
The rest of the potential remains the same. Thereby the field $\phi$ can
transform as any one-dimensional representation of
\Groupname{D}{4}. $\lambda _{3}$ and $\tilde{\lambda} _{3}$ lead to
$\beta= \gamma$ such that $\alpha = \beta = \gamma$ is enforced. The
accidental $U(1)$ can be restored, if $\tilde{\lambda} _{3} = - \lambda_{3}$
is chosen, since then \Eqref{Higgs3D4} simplifies to $- 4 \,
\lambda_{3} |\psi _{1} ^{\dagger} \, \psi_{2}|^2$.

\noindent Since \Groupname{D}{6} has also been mentioned as flavor
symmetry in the literature and is the next smallest \Groupname{D}{n}
symmetry after \Groupname{D}{5}, we briefly comment on
\Groupname{D}{6} invariant three Higgs potentials. If the three fields
transform as faithful two-dimensional and as trivial
representation, their potential incorporates an accidental $U(1)$
symmetry. However, using instead one of the two
further one-dimensional representations of \Groupname{D}{6} which are not
present in \Groupname{D}{5} one can get rid of this $U(1)$. Products of the faithful
representation with these have the structure $\Rep{1} \times \Rep{2} =
\Rep{2} ^{\prime}$ and therefore lead together with $\Rep{2} \times
\Rep{2} = \MoreRep{1}{1} + \Rep{1} ^{\prime} + \Rep{2} ^{\prime}$ to a
potential which coincides with the one obtained from \Groupname{D}{3}.

\noindent This demonstrates that a thorough discussion of the Higgs
potential is always necessary to ensure the validity of the model as a
whole. A more complete discussion about the possible potentials arising
from \Groupname{D}{n} flavor symmetries and also \Doub{D}{n}
symmetries will be given elsewhere \cite{comingsoon}.

\subsection{Four Higgs Potential}
\label{sec:Higgs4}

In this Subsection we consider a potential containing four Higgs fields. There exist
two possible choices. First we can augment our three Higgs potential
with a further Higgs field $\chi$ transforming as one-dimensional
representation. If $\phi \sim \MoreRep{1}{i}$ then $\chi$ should
transform as $\MoreRep{1}{j}$ with $\rm i \neq j$. Writing down all
possible \Groupname{D}{5} invariant couplings shows that they cannot
break the $U(1) _{X}$ symmetry. Therefore we will consider a four
Higgs potential with fields $\chi_{i}$ and $\psi _{i}$, $i=1,2$. Each
pair forms a doublet under \Groupname{D}{5}, without loss of
generality: $\left( \begin{array}{c} \chi _{1} \\ \chi_{2} \end{array}
\right) \sim \MoreRep{2}{1}$ and $\left( \begin{array}{c} \psi _{1} \\ \psi_{2} \end{array}
\right) \sim \MoreRep{2}{2}$. The potential then has the following
form:
\small
\begin{eqnarray}\nonumber
V_{4} (\chi _{i}, \psi_{i}) &=& -\mu _{1}^{2} \, \sum \limits _{i=1}
^{2} \chi ^{\dagger} _{i} \, \chi _{i}-\mu _{2} ^{2} \sum \limits
_{i=1} ^{2} \psi ^{\dagger} _{i} \, \psi _{i} +  \lambda _{1} \, \left( \sum \limits _{i=1}
^{2} \chi ^{\dagger} _{i} \, \chi _{i} \right)^{2} +
\tilde{\lambda}_{1} \, \left( \sum \limits
_{i=1} ^{2} \psi ^{\dagger} _{i} \, \psi _{i} \right) ^{2}\\ \nonumber
&+& \lambda _{2} \, \left( \chi ^{\dagger} _{1} \, \chi _{1}-\chi
  ^{\dagger} _{2} \, \chi _{2} \right) ^{2} + \lambda _{3} \, |\chi
^{\dagger} _{1} \, \chi _{2} |^{2} +  \tilde{\lambda} _{2} \, \left( \psi ^{\dagger} _{1} \, \psi _{1}-\psi
  ^{\dagger} _{2} \, \psi _{2} \right) ^{2} + \tilde{\lambda} _{3} \, |\psi
^{\dagger} _{1} \, \psi _{2} |^{2} \\ \nonumber
&+& \sigma _{1} \, \left( \sum \limits _{i=1}
^{2} \chi ^{\dagger} _{i} \, \chi _{i} \right) \,  \left( \sum \limits _{j=1}
^{2} \psi ^{\dagger} _{j} \, \psi _{j} \right) + \sigma_{2} \,
\left(\chi ^{\dagger} _{1} \, \chi _{1}-\chi ^{\dagger} _{2} \, \chi
  _{2} \right) \, \left(\psi ^{\dagger} _{1} \, \psi _{1}-\psi ^{\dagger} _{2} \, \psi
  _{2} \right)  \\ \nonumber
&+& \left\{ \tau _{1} \, \left( \chi ^{\dagger} _{1} \, \psi _{1}
  \right) \, \left( \chi ^{\dagger} _{2} \, \psi _{2} \right) +
  \mathrm{h.c.} \right\} +  \left\{ \tau _{2} \, \left( \chi ^{\dagger} _{1} \, \psi _{2}
  \right) \, \left( \chi ^{\dagger} _{2} \, \psi _{1} \right) +
  \mathrm{h.c.} \right\}  \\ \nonumber
&+& \left\{ \kappa _{1} \, \left[ \left( \chi ^{\dagger} _{1} \, \chi
      _{2} \right) \, \left( \chi ^{\dagger} _{1} \, \psi _{2} \right)
    + \left( \chi ^{\dagger} _{2} \, \chi _{1} \right) \, \left( \chi
      ^{\dagger} _{2} \, \psi _{1} \right)
  \right] + \mathrm{h.c.} \right\} \\ \nonumber 
&+& \left\{ \kappa _{2} \, \left[ \left( \psi ^{\dagger} _{1} \, \psi
      _{2} \right) \, \left( \chi ^{\dagger} _{2} \, \psi _{2} \right)
    + \left( \psi ^{\dagger} _{2} \, \psi _{1} \right) \, \left( \chi
      ^{\dagger} _{1} \, \psi _{1} \right)
  \right] + \mathrm{h.c.} \right\} \\ 
&+& \kappa _{3} \, \left[ | \chi ^{\dagger} _{1} \, \psi _{1} |^{2}  +
|\chi ^{\dagger} _{2} \, \psi _{2} | ^{2} \right] +  \kappa _{4} \, \left[ | \chi ^{\dagger} _{1} \, \psi _{2} |^{2}  +
|\chi ^{\dagger} _{2} \, \psi _{1} | ^{2} \right]
\end{eqnarray}
\normalsize
where the couplings $\tau_{1,2}$ and $\kappa _{1,2}$ are in general complex. We
checked that this potential does not have any accidental (global)
symmetries. Assuming that the fields $\chi_{1,2}$, $\psi _{1,2}$
transform in the following way:
\begin{equation}\nonumber
\chi_{1} \;\;\; \rightarrow \;\;\; \chi_{1} \; \mathrm{e} ^{i \, \alpha}
\; , \;\; \chi_{2} \;\;\; \rightarrow \;\;\; \chi_{2} \; \mathrm{e} ^{i
  \, \beta} \; , \;\; \psi_{1} \;\;\; \rightarrow \;\;\; \psi_{1} \; \mathrm{e} ^{i \, \gamma}
\; , \;\; \psi_{2} \;\;\; \rightarrow \;\;\; \psi_{2} \; \mathrm{e} ^{i
  \, \delta} \; ,
\end{equation}
one finds that the couplings $\mu_{1,2}, \lambda_{1,2,3},
\tilde{\lambda} _{1,2,3}, \sigma_{1,2}, \kappa_{3,4}$ leave the full $U(1) ^{4}$ invariant,
$\tau_{1,2}$ breaks it down to $U(1) ^{3}$ and $\kappa_{1,2}$ down to $U(1) ^{2}$, i.e. none
of the couplings itself is only invariant under $U(1)
_{Y}$. $\tau_{1,2}$ leave the same $U(1) ^{3}$ invariant with the
condition $\alpha= \gamma + \delta - \beta$. The $U(1)
^{2}$ symmetries which are preserved by $\kappa_{1,2}$ are constrained
by the conditions $2 \, \alpha= \beta + \delta$, $2 \, \beta = \alpha
+ \gamma$ and $2 \, \delta = \beta + \gamma$, $2 \, \gamma= \alpha +
\delta$, respectively. As one can see only $\kappa_{1} \neq 0$ and
$\kappa_{2} \neq 0$ can reduce $U(1)^{4}$ to $U(1)_{Y}$, i.e. taking
the $\tau_{1,2}$ terms with only the $\kappa_{1}$ term still leaves
the potential invariant under $U(1) ^{2}$. Consequently, none of the VEV
conditions should enforce $\kappa_{1}$ or $\kappa_{2}$ to vanish. A
simple example for this is the configuration $\langle \chi_{1} \rangle
\neq 0$, $\langle \psi_{2} \rangle \neq 0$ and $\langle \chi_{2}
\rangle = \langle \psi_{1} \rangle = 0$ with all VEVs being real. It leads to
$\kappa_{1} =0$. However, it cannot produce phenomenological viable
mass matrices anyway as discussed above.

\noindent In the following we show that the VEV configuration which is
used in the zeroth order approximation in our numerical study represents one
possible minimum of the Higgs potential $V_{4}$. As one can see, the
equivalence of all four VEVs is not obligatory, since for example
$\mu_{1}$ and $\mu_{2}$ and $\lambda_{1}$ and $\tilde{\lambda}_{1}$
are not restricted to have the same value, respectively. Therefore we
search for a symmetry which can maintain these restrictions such that
the equivalence of all four VEVs becomes more natural. The simplest
choice is to first interchange the fields $\chi_{i}$ with
$\psi_{i}$ in order to enforce for example the equivalence of
$\mu_{1}$ and $\mu_{2}$ and to further exchange the fields $\chi_{1}$ and
$\chi _{2}$ preventing the couplings $\kappa_{1,2}$ from
being set to zero \footnote{The exchange of the fields $\psi_{1}$ and $\psi_{2}$
gives the same result.}. This symmetry will be called $T$ in the following. It
restricts the parameters as follows:
\begin{equation}
\mu_{1} = \mu_{2} \; , \;\; \lambda_{i} = \tilde{\lambda}_{i} \; , \;\;
\sigma_{2} = 0 \; , \;\; \tau_{1} = \tau_{2} ^{\star} \; , \;\; \kappa_{1} = \kappa_{2} ^{\star}
\; , \;\; \kappa_{3}= \kappa_{4} \; .
\end{equation}
Note that setting $\sigma_{2}$ to zero does not lead to an accidental
continuous symmetry. Especially, we do not enforce $\kappa_{1,2}$ to
vanish. Note also that changing the order of the actions $\chi_{i}
\; \leftrightarrow  \; \psi_{i}$ and $\chi_{1} \; \leftrightarrow \;
\chi_{2}$ does not change the result. 

\noindent Next we analyze the potential
invariant under $\mbox{\Groupname{D}{5}} \, \times \, T$ for real VEVs $\langle \chi_{1} \rangle = \frac{v}{\sqrt{2}} \, \cos
(\alpha)$, $\langle \chi_{2} \rangle = \frac{v}{\sqrt{2}} \, \sin (\alpha)$, $\langle
\psi_{1} \rangle = \frac{u}{\sqrt{2}} \, \cos (\beta)$ and $\langle \psi_{2} \rangle = \frac{u}{\sqrt{2}}
\, \sin (\beta)$. The form of the potential at the extremum is:
\small
\begin{eqnarray}\nonumber
V_{4 \, T \; min} &=& - \frac{1}{2} \, \mu_{1} ^{2} \, (u^2 + v^2) + \frac{1}{32} \, (u^4 + v^4) \, \left( 8 \,
  \lambda_{1} + 4 \, \lambda_{2} + \lambda_{3} \right) + \frac{1}{4}
\, u^2 \, v^2 \, (\sigma_{1} + \kappa_{3})\\ \nonumber &+& \frac{1}{32} \, (v^4 \,
\cos (4 \, \alpha) + u^4 \, \cos( 4 \, \beta)) \, \left( 4 \,
  \lambda_{2} - \lambda_{3} \right) + \frac{1}{4} \, u \, v \, \left[
  u^2 \, \cos (\alpha - \beta) \, \sin (2 \, \beta) \right. \\
 &+& \left. v^2 \, \sin (2 \,\alpha) \, \sin (\alpha + \beta) \right] \, \mathrm{Re}
(\kappa_{1}) + \frac{1}{4} \, u^2 \, v^2 \, \sin (2 \, \alpha) \, \sin
(2 \, \beta) \, \mathrm{Re} (\tau_{1})
\end{eqnarray}
\normalsize
The minimization conditions which can be deduced from $V _{4 \, T
  \; min}$ are:
\footnotesize
\begin{subequations}
\begin{eqnarray}\label{mincond1}
\frac{\partial V _{4 \, T \; min}}{\partial \alpha} &=& -\frac{1}{8} \,
v^4 \, \sin (4 \, \alpha) \, y + \frac{1}{2} \, u^2 \, v^2 \, \cos (2
\, \alpha) \, \sin (2 \, \beta) \, \mathrm{Re} (\tau_{1}) \\ \nonumber
&+& \frac{1}{4} \, u \, v \, \left[ v^2 \, \left( \cos (2 \, \alpha)
    \, \sin (\alpha + \beta) + \sin (3 \, \alpha + \beta) \right) -
  u^2 \, \sin (\alpha - \beta) \, \sin (2 \, \beta) \right] \,
\mathrm{Re} (\kappa_{1})
\end{eqnarray}
\begin{eqnarray}\label{mincond2}
\frac{\partial V _{4 \, T \; min}}{\partial \beta} &=& - \frac{1}{8} \,
u^4 \, \sin (4 \, \beta) \, y + \frac{1}{2} \, u^2 \, v^2 \, \sin (2
\, \alpha) \, \cos (2 \, \beta) \, \mathrm{Re} (\tau_{1})\\ \nonumber
&+& \frac{1}{4} \, u \, v \, \left[ u^2 \, \left( \cos (\alpha -
    \beta) \, \cos(2 \, \beta) + \cos (\alpha - 3 \, \beta) \right) +
  v^2 \, \sin (2 \, \alpha) \, \cos (\alpha + \beta) \right] \,
\mathrm{Re} (\kappa_{1})
\end{eqnarray}
\end{subequations}
\normalsize
where $y= 4 \, \lambda_{2} - \lambda_{3}$. \Eqref{mincond1} and \Eqref{mincond2} are
fulfilled for $\alpha = \frac{\pi}{4}$ and $\beta=
\frac{\pi}{4}$. Then each of the terms vanishes separately, especially
there is no constraint on $\mathrm{Re} (\tau _{1})$, $\mathrm{Re}
(\kappa_{1})$ or $4 \, \lambda_{2} - \lambda_{3}$. This is important,
since constraining these
parameters to be zero could lead to accidental symmetries. For $\alpha= \frac{\pi}{4}$ and
$\beta= \frac{\pi}{4}$ there is also one solution with
$u=v$. Therefore the equivalence of all (real) VEVs is a natural
result of the potential.

\noindent Apart from this zeroth order solution it is important to check
whether phenomenological viable VEV configurations can be a minimum of
the potential for an appropriate choice of parameters. As we tried to
restrict ourselves above to SCPV, it is especially necessary to find out
whether this is possible for the chosen VEVs in our numerical
examples. Unfortunately, it turns out to be impossible for the
potential invariant under $\mbox{\Groupname{D}{5}} \, \times \,
T$. For VEVs parameterized as $\langle \chi_{1} \rangle =
\frac{v_{1}}{\sqrt{2}} \, \mathrm{e}^{i \, \alpha}$, $\langle \chi_{2} \rangle
= \frac{v_{2}}{\sqrt{2}} \, \mathrm{e}^{i \, \beta}$, $\langle \psi_{1}
\rangle = \frac{v_{3}}{\sqrt{2}} \, \mathrm{e}^{i \, \gamma}$ and $\langle \psi_{2}
\rangle= \frac{v_{4}}{\sqrt{2}}$ one can deduce, for example, the following
equations from the minimization conditions for $v_{i} \neq 0$, $\alpha
\neq 0$, $\beta \neq 0$, $\gamma \neq 0$:
\begin{subequations}
\begin{gather}
5 \, v_3 \, v_4 \, \mathrm{Re} (\kappa_{1}) \, (v_1 \, v_3 \, \sin
(\alpha -2 \, \gamma) - v_2 \, v_4 \, \sin (\beta + \gamma)) =0\\
5 \, v_1 \, v_2 \, \mathrm{Re} (\kappa_{1}) \, (v_1 \, v_4 \, \sin (2
\, \alpha - \beta) + v_2 \, v_3 \, \sin (\alpha - 2 \, \beta + \gamma))=0
\end{gather}
\end{subequations}
These directly lead to the conclusion that $\mathrm{Re}
(\kappa_{1})=0$. As we consider SCPV, also $\mathrm{Im}
(\kappa_{1})=0$ and therefore the coupling $\kappa_{1}$
vanishes \footnote{Actually in general even more parameters of the potential
are constrained to be zero or have to fulfill certain relations.}. This
increases the symmetry of the potential, as explained above. With
$v_{i} \neq 0$, $\alpha \neq 0$, $\beta \neq 0$, $\gamma \neq 0$ it is
then clear that this additional symmetry will be broken and hence
further massless Goldstone bosons will appear which are
phenomenologically unacceptable. In this case we have not gained
anything by discussing the four Higgs potential compared to the three
Higgs one. Abandoning the $T$ symmetry and only requiring that the
potential is invariant under \Groupname{D}{5} does not change the
situation, since then one can deduce the equations:
\begin{subequations}
\begin{gather}
5 \, v_3 \, v_4 \, \mathrm{Re} (\kappa_{2}) \, (v_1 \, v_3 \,\sin
(\alpha - 2 \,\gamma) - v_2 \, v_4 \, \sin (\beta +\gamma)) =0\\
5 \, v_1 \, v_2 \, \mathrm{Re} (\kappa_{1}) \, (v_1 \, v_4 \, \sin (2
\, \alpha - \beta) + v_2 \, v_3 \, \sin (\alpha - 2 \,\beta + \gamma))=0
\end{gather}
\end{subequations}
These enforce the vanishing of $\mathrm{Re}(\kappa_{1})$ and
$\mathrm{Re} (\kappa_{2})$ for general VEV configurations. Again
$\mathrm{Im}(\kappa_{1,2})$ are already set to zero, since we want to
study the case of SCPV. In the end, the constraints $\kappa_{1}=0$ and
$\kappa_{2}=0$ lead to an increase of the symmetry of the
potential. Similar to the case above this further symmetry is broken
by arbitrary VEV configurations resulting in extra Goldstone
bosons. This proves that SCPV can only exist for special VEV
configurations, but not in general. \\
\noindent For a general \Groupname{D}{5} invariant
four Higgs potential with complex parameters one can successfully solve
all minimization conditions without the necessity to set parameters
to zero. Furthermore one is able to maintain that all masses of the
Higgs fields at this extremum are positive, i.e. this extremum can be
a minimum of the potential. As all relevant equations are invariant
under $v_{i} \rightarrow - v_{i}$, the VEV configurations
$\langle \chi_{1} \rangle= \frac{v_{1}}{\sqrt{2}} \, \mathrm{e}^{i \,
  \alpha}$, $\langle \chi_{2} \rangle= \frac{v_{2}}{\sqrt{2}} \, \mathrm{e}^{i
\, \beta}$, $\langle \psi_{1} \rangle= \frac{v_{3}}{\sqrt{2}} \, \mathrm{e}^{i
\, \gamma}$, $\langle \psi_{2} \rangle= \frac{v_{4}}{\sqrt{2}}$ and $\langle \chi_{1} \rangle= \frac{-v_{1}}{\sqrt{2}} \, \mathrm{e}^{i \,
  \alpha}$, $\langle \chi_{2} \rangle= \frac{-v_{2}}{\sqrt{2}} \, \mathrm{e}^{i
\, \beta}$, $\langle \psi_{1} \rangle= \frac{-v_{3}}{\sqrt{2}} \, \mathrm{e}^{i
\, \gamma}$, $\langle \psi_{2} \rangle= \frac{-v_{4}}{\sqrt{2}}$ are
degenerate. Finally, one can check numerically whether the potential
is stable as a whole. This clearly is not a proof of the stability of
the potential, but is enough for our considerations.\\ 
\noindent All this has been
done for the two VEV configurations used in the numerical
examples. The parameters of the potential can be chosen in such a way
that all constraints are fulfilled. The mass of the lightest Higgs
field is usually smaller ($\sim 40 \GeV$) than the
experimental bounds ($\leq 114.4 \GeV$) \cite{lepbound}, if the mass parameters
$\mu_{i}$ are of the order of the electroweak scale $(100 - 200 \GeV)$
and the quartic couplings are in the perturbative range. This problem
can be cured by simply assuming that the $\mu_{i}$s are larger than
$\mathcal{O}(100 \GeV)$ or adding some other mass dimension two terms
which break \Groupname{D}{5}. In order to pass not only the direct Higgs mass
bounds, but also the stringent bounds on FCNCs, the Higgs masses
should be even larger
than a few $\TeV$. The mechanism of
adding \Groupname{D}{5} breaking terms is unmotivated from the
theoretical point of view, but seems to be necessary for a
phenomenological viable model in this context.\\ 
\noindent One could ask whether
it is also possible to achieve that arbitrary VEV configurations can be
minima of the potential, if this is invariant under $\mbox{\Groupname{D}{5}}
\, \times \, T$. The
answer is no, since one can deduce three linear independent equations
containing $\mathrm{Re}(\kappa_{1})$, $\mathrm{Im}(\kappa_{1})$ and
$\mathrm{Re}(\tau_{1})$ which are in general only solved, if
$\mathrm{Re}(\kappa_{1})=0$, $\mathrm{Im}(\kappa_{1})=0$ and
$\mathrm{Re}(\tau_{1})=0$. Again, the minimization conditions enforce
a parameter setup which leads to an additional global symmetry in the
Higgs potential.\\
\noindent Finally, we compare the \Groupname{D}{5} invariant potential
of four Higgs fields to the equivalent one in
\Groupname{D}{6}. Similar to \Groupname{D}{5} also \Groupname{D}{6}
has two inequivalent two-dimensional representations (one faithful and
one unfaithful one). However, in contrast to \Groupname{D}{5} the \Groupname{D}{6}
invariant four Higgs potential contains a further $U(1)$ symmetry. The
reason for this is the \Groupname{D}{6} product structure $\MoreRep{2}{i} \times
\MoreRep{2}{i} = \MoreRep{1}{1} + \MoreRep{1}{4} + \MoreRep{2}{2}$ for
$\rm i=1,2$ and $\MoreRep{2}{1} \times \MoreRep{2}{2} = \MoreRep{1}{2}
+ \MoreRep{1}{3} + \MoreRep{2}{1}$ which does not allow for invariant
couplings of the form $\MoreRep{2}{i} ^{3} \, \MoreRep{2}{j}$ with $\rm
i \neq j$. Precisely, these couplings, $\kappa_{1}$ and $\kappa_{2}$,
exist in \Groupname{D}{5} and
therefore prevent the potential from having an accidental $U(1)$.

\section{Extensions of the Model}
\label{sec:extensions}

\noindent Finally, we would like to comment on how the model has to be
changed in order to be embedded into an $SO(10)$ GUT and - maybe
simultaneously - into a continuous flavor symmetry, like $SO(3) _{f}$ or
$SU(3) _{f}$. This is desirable, since GUTs turned out to be very
successful in unifying the SM gauge interactions and fermions of one
generation and in explaining, for example, charge quantization. These
features should not be given up when flavored models are
considered. Second, the embedding of a discrete flavor symmetry into a
continuous group $G_{f}$ allows one to unify it with the GUT group
being also continuous into one group containing gauge and flavor
symmetries. Attempts to find such a group can be found in the
literature \cite{GUTflavor}. Albeit these have not been very
successful, the idea is still appealing. Furthermore gauged symmetries
are the only ones which remain unbroken in the presence of quantum
gravitational corrections \cite{quantumgrav} which suggests that any flavor symmetry should also
be gauged. However, gauging a discrete symmetry can be performed in
the easiest way, if it is embedded into a continuous one which is then
gauged. Nevertheless, in the context of string theory discrete flavor
symmetries could also arise without such an embedding.

\noindent Since all fermions of one generation reside in the
$\Rep{16}$ of $SO(10)$ they need to transform in the same way under
\Groupname{D}{5}, for example as $\MoreRep{1}{1} + \MoreRep{2}{1}$. In
our minimal model with just the four Higgs fields
$\chi_{i}$ and $\psi_{i}$ the resulting mass matrices do hardly lead
  to phenomenological viable masses at tree-level and at low
  energies. Therefore we have to extend the Higgs sector by at least
  one Higgs field $\phi$ transforming trivially under
  \Groupname{D}{5}. The mass matrices are then of the form\\
\small
\rule[0in]{-0.5in}{0in}\parbox{7in}{\begin{equation}\label{SO10Diracmasses}
\mathcal{M} _{u, \nu} = \left( \begin{array}{ccc} 
    \alpha ^{u , \nu} _{0} \, \langle \phi \rangle ^{\star}  &  \alpha
    ^{u,\nu} _{1} \, \langle  \chi_{1} \rangle ^{\star} &
    \alpha ^{u , \nu} _{1} \, \langle  \chi_{2} \rangle ^{\star} \\
    \alpha ^{u , \nu} _{2} \, \langle  \chi_{1} \rangle ^{\star} &
    \alpha ^{u , \nu} _{4} \, \langle  \psi_{1} \rangle ^{\star} &
    \alpha ^{u , \nu} _{3} \, \langle  \phi \rangle ^{\star} \\
    \alpha ^{u , \nu} _{2} \, \langle  \chi_{2} \rangle  ^{\star} & 
    \alpha ^{u , \nu} _{3} \, \langle \phi \rangle ^{\star} &
    \alpha ^{u , \nu} _{4} \, \langle \psi_{2} \rangle ^{\star}
    \end{array}
    \right) \; , \; \mathcal{M} _{d, l} = \left( \begin{array}{ccc} 
     \alpha ^{d,l} _{0} \, \langle \phi \rangle & \alpha ^{d,l} _{1}
     \, \langle \chi _{2} \rangle & \alpha
    ^{d,l} _{1} \, \langle \chi_{1} \rangle  \\
    \alpha ^{d,l} _{2} \, \langle  \chi _{2} \rangle & \alpha ^{d,l}
    _{4}  \, \langle  \psi _{2} \rangle &  \alpha
    ^{d,l} _{3} \, \langle \phi \rangle  \\
    \alpha ^{d,l} _{2} \, \langle \chi_{1} \rangle &  \alpha ^{d,l}
    _{3} \, \langle \phi \rangle & \alpha ^{d,l} _{4} \, \langle \psi_{1} \rangle 
    \end{array}
    \right) \; ,
\end{equation}}\\
\normalsize
i.e. the Higgs field $\phi$ fills the zeros in the $(1,1)$,$(2,3)$ and
$(3,2)$ elements. Note that the form of the right-handed Majorana mass
terms does not change.
In a complete $SO(10)$ model the Higgs doublet fields have to be
embedded into the representations $\Rep{10}$, $\Rep{120}$ and
$\overline{\Rep{126}}$, since these do couple to $\Rep{16} \times
\Rep{16}$. Still this setup has to be embedded into the continuous
flavor group $G_{f}$. For $G_{f}$ being $SO(3) _{f}$ this is not
possible, since we cannot identify $\MoreRep{1}{1} + \MoreRep{2}{i}$
with the fundamental representation of $SO(3) _{f}$. The same holds
for $SU(3) _{f}$. In order to do so, the first generation has to
transform as $\MoreRep{1}{2}$ rather than $\MoreRep{1}{1}$. This leads
to a sign in the $(1,3)$ and $(3,1)$ elements of the mass matrices in
\Eqref{SO10Diracmasses}, but does not alter the discussion. The five
Higgs fields
$\chi_{i}$, $\psi_{i}$ and $\phi$ $\sim \MoreRep{1}{1} +
\MoreRep{2}{1} + \MoreRep{2}{2}$ can be identified with the $\Rep{5}$
of $SO(3) _{f}$ and together with an additional field $\phi^{\prime}
\sim \MoreRep{1}{1}$ also with the six-dimensional representation of
$SU(3) _{f}$. 

\noindent A more minimal choice for an embedding into $SO(10) \times
G_{f}$ would be given by the three generations transforming as
$\MoreRep{1}{2} + \MoreRep{2}{1}$ and three Higgs fields doing the
same. Unfortunately, this leads to traceless mass matrices for the
fermions which seem to be highly disfavored by the observed mass
hierarchies among the generations. This problem can be cured by adding
another Higgs field transforming trivially under
\Groupname{D}{5}. Furthermore this increases the number of allowed
Yukawa couplings by two. Since the added Higgs field transforms as
$\MoreRep{1}{1}$, the model can still be embedded into the
continuous flavor symmetries $SO(3)_{f}$ and $SU(3)_{f}$ with this
field being identified with the singlet of $SO(3)_{f}$ or
$SU(3)_{f}$. Although we showed that the Higgs sector is not phenomenological
viable in this case (see \Secref{sec:Higgspotentials}), we cannot exclude
it as a GUT model, because the
Higgs couplings might change through the embedding of the $SU(2) _{L}$
Higgs doublet fields into $SO(10)$ representations.

\section{Conclusions and Outlook}
\label{sec:conclusion}

In this paper, we constructed a minimal model with the SM gauge group enlarged
by the flavor symmetry \Groupname{D}{5}. Both are broken only
spontaneously at the electroweak scale. We chose \Groupname{D}{5},
since it is the smallest discrete group with two inequivalent irreducible
two-dimensional representations. We demanded the left- and
left-handed conjugate
fields of the three generations to unify partially, i.e. transform
as $\Rep{1} + \Rep{2}$ under \Groupname{D}{5}, combined with the
requirement that our model should be embeddable at least into the
Pati-Salam gauge group. Furthermore we have chosen the minimal
possible number of Higgs doublets with a potential free of accidental
symmetries and did not include scalar
fields transforming as $SU(2) _{L}$ triplets or gauge singlets. We
showed that under these constraints a minimal model can be built in
which the left-handed fields transform as $\MoreRep{1}{1} +
\MoreRep{2}{2}$ under \Groupname{D}{5}, the left-handed conjugate ones
as $\MoreRep{1}{1} + \MoreRep{2}{1}$ and the four Higgses $\chi_{i}$ and
$\psi_{i}$ ($i=1,2$) as $\MoreRep{2}{1} + \MoreRep{2}{2}$. By a
numerical study we showed that all fermion masses and mixing
parameters can be accommodated at tree-level. We considered the case of Majorana as
well as Dirac neutrinos and we discussed the results of the unmeasured
leptonic quantities. By our choice the spectrum of the light neutrinos
is always normally ordered. The structure of the right-handed neutrino
mass matrix is (almost) trivial, since we did not include SM gauge
singlets. As a consequence two of the right-handed neutrinos are
degenerate at tree-level. We compared the structure of the
\Groupname{D}{5} invariant mass matrices with those of
\Groupname{D}{3} invariant ones which are often discussed in the
literature. The main difference is the tendency to get more texture
zeros for a similar assignment of fermions and Higgs fields
arising from the existence of the two inequivalent
two-dimensional representations in \Groupname{D}{5}. We then turned to
a discussion of the Higgs sector and found that all potentials with
three Higgs fields transforming as $\Rep{1} + \Rep{2}$ are not only
\Groupname{D}{5} invariant, but also incorporate an accidental $U(1)$
symmetry which is broken by any VEV
configuration leading to phenomenological viable mass matrices for the
fermions at tree-level. To find the group theoretical reason for this
accidental $U(1)$ we considered similar potentials invariant under
\Groupname{D}{3} and \Groupname{D}{4}, respectively, and found that
they do not have an accidental $U(1)$ symmetry. The difference lies in
the \Groupname{D}{5} product structure $\MoreRep{1}{1,2} \times \Rep{2} =
\Rep{2}$ and $\Rep{2} \times \Rep{2} = \MoreRep{1}{1} + \MoreRep{1}{2}
+ \Rep{2} ^{\prime}$ such that the coupling $\Rep{2} \times \Rep{2}
\times \Rep{2} \times \MoreRep{1}{1,2}$ is not invariant under
\Groupname{D}{5}. Therefore we had to extend the Higgs sector to four
fields $\chi_{i}$ and $\psi_{i}$ transforming as the doublets of
\Groupname{D}{5}. We explicitly showed that this potential is free of
accidental symmetries and analyzed its VEV configurations. For a
zeroth order solution we
imposed a further discrete symmetry - called $T$ - on the potential in
order to maintain the configuration
that all (real) VEVs are equal as natural outcome of the minimization
conditions. In a second step we proved that SCPV is not possible for
the VEV configurations used in our numerical examples of the fermion
mass matrices. Nevertheless these configurations can be minima of the
\Groupname{D}{5} invariant four Higgs potential, if its parameters are
complex. Furthermore we calculated the masses for the Higgs fields. We
found that they are naturally of the order $\mathcal{O}(100 \GeV)$
up to $\mathcal{O}(1 \TeV)$ with the smallest mass below the LEP bound of
$114.4 \GeV$ \cite{lepbound}, if the mass parameters of the potential are of the order
of the electroweak scale and the quartic couplings are in the
perturbative regime. Therefore FCNCs might be a problem which can
probably be cured by adding large mass dimension two terms which break
\Groupname{D}{5}. In our numerical examples
the FCNCs involving the first generation get additionally suppressed,
since the relevant Yukawa
couplings are at most $10^{-4}$. Finally, we considered extensions of
our low energy model and showed the necessary changes
in the particle assignment and content to achieve the embedding into
$SO(10) \times G_{f}$ where $G_{f}$ can be either $SO(3)_{f}$ or $SU(3)_{f}$.

\noindent In our numerical examples the VEVs of the fields $\chi_{i}$
and $\psi_{i}$ break \Groupname{D}{5} completely. However,
\Groupname{D}{5} has two non-trivial abelian subgroups
\Groupname{Z}{2} and \Groupname{Z}{5} which can be generated by the
generator $\rm B$ and the generator $\rm A$ alone, respectively. As
one can see, \Groupname{Z}{5} is always broken by a non-vanishing VEV
of $\chi_{i}$ and $\psi_{i}$. In contrast to this a residual
\Groupname{Z}{2} is preserved in the Lagrangian, if $\langle \chi_{1}
\rangle = \langle \chi_{2} \rangle$ and $\langle \psi_{1} \rangle =
\langle \psi_{2} \rangle$. Interestingly, the resulting mass matrices $\mathcal{M}
_{u,\nu}$ and $\mathcal{M} _{d,l}$ are then invariant under the
interchange of the second and third generation and therefore produce a
maximal mixing in the 2-3 sector, a vanishing mixing in the 1-3 sector
and leave the mixing angle $\theta_{12}$ undetermined. For an exact
\Groupname{Z}{2} thus the mixing matrices $V_{CKM}$ and $U_{MNS}$ have
two vanishing mixing angles $\theta_{13}$ and $\theta_{23}$, since the
maximal mixing angles in the 2-3 sectors of the up-type quarks
(neutrinos) and the down-type
quarks (charged leptons) cancel each other. This
leads to the conclusion that this residual \Groupname{Z}{2} is only weakly
broken in the quark sector, but strongly broken in the lepton/neutrino
sector. Actually the equalities $\langle \chi_{1} \rangle = \langle \chi_{2} \rangle$
and $\langle \psi_{1} \rangle = \langle \psi_{2} \rangle$ have also
been employed when we searched for an appropriate zeroth order
structure of the fermion mass matrices in our phenomenological
analysis (see \Secref{sec:numerics}). By considering the non-trivial
subgroups of \Groupname{D}{5} this choice gains further significance.

\noindent This discussion can be compared with the studies of the
non-trivial subgroups of \Groupname{A}{4}
\cite{A4subgroups}. \Groupname{A}{4} can be broken to either
\Groupname{Z}{2} or \Groupname{Z}{3} by different VEV configurations
of Higgs fields forming a triplet under \Groupname{A}{4}. It turns out
that preserving the subgroup \Groupname{Z}{3} for charged fermions
leads to $V_{CKM}= \mathbb{1}$ whereas \Groupname{Z}{2} is preserved
in the neutrino sector leading to tri-bi-maximal mixing. Concerning the
quark sector our flavor symmetry \Groupname{D}{5} has the advantage
that its breaking to \Groupname{Z}{2} leads to vanishing 1-3 and 2-3
mixing, but does not constrain the Cabibbo angle. In this way we can
explain why the Cabibbo angle is about one order of magnitude larger
than the two other mixing angles whereas models using \Groupname{A}{4}
might have problems to generate a 1-2 mixing angle being large enough to
accommodate the data. On the other hand in our minimal model shown
here the residual subgroups of
\Groupname{D}{5} can hardly give reason for the tri-bi-maximal or bi-maximal
mixing pattern observed in the leptonic sector which can nicely be
explained by the residual \Groupname{Z}{2} symmetry of
\Groupname{A}{4} in the neutrino sector.

\vspace{0.5cm}
\begin{center}
{\bf Acknowledgments}
\end{center}
This work was supported by the ``Deutsche Forschungsgemeinschaft'' in the 
``Sonderforschungsbereich 375 f\"ur Astroteilchenphysik''.

\small
\appendix

\section{Details of Group Theory}
\label{app:grouptheory}

\noindent Here, we show the Clebsch Gordan coefficients for all
Kronecker products in case that none of the representations is complex
conjugated. This choice corresponds to the Yukawa couplings for the
down-type quarks and charged leptons (see \Secref{sec:model}). All
other Clebsch Gordan coefficients needed for example for the quartic
couplings in the Higgs sector and the Yukawas for the up-type quarks
which involve at least one complex conjugated representation can be
generated from the given Clebsch Gordan coefficients taking into
account the similarity transformation between the representation
matrices and its complex conjugates as shown in \Secref{sec:grouptheorydn}.

\noindent For $A \sim \MoreRep{1}{i}$ and $B \sim \MoreRep{1}{j}$ the product is
$A \, B \sim \MoreRep{1}{\scriptsize \rm
  (i+j) mod 2 +1 \normalsize}$. Combining the two-dimensional
representation $\left( \begin{array}{c} a_{1} \\ a_{2} \end{array}
\right) \sim \MoreRep{2}{i}$ with the trivial singlet $A \sim
\MoreRep{1}{1}$ leads to  $\left( \begin{array}{c} A \, a_{1} \\ A \, a_{2} \end{array}
\right) \sim \MoreRep{2}{i}$. Similarly for the non-trivial singlet
$B \sim \MoreRep{1}{2}$ one finds  $\left( \begin{array}{c} B \, a_{1}
    \\-B \, a_{2} \end{array}
\right) \sim \MoreRep{2}{i}$.

\noindent The \Groupname{D}{5} covariant combinations of $\MoreRep{2}{1} \times
\MoreRep{2}{1}$  for $\left( \begin{array}{c} a_{1} \\ a_{2} \end{array} \right) \, , \,
  \left( \begin{array}{c} a ^{\prime}_{1} \\ a ^{\prime} _{2}
      \end{array} \right) \sim \MoreRep{2}{1}$ are $a_{1} \, a_{2} ^{\prime} + a_{2} \, a^{\prime} _{1} \sim
\MoreRep{1}{1}$, $a_{1} \, a_{2} ^{\prime}-a_{2} \, a^{\prime} _{1}
\sim \MoreRep{1}{2}$ and $\left( \begin{array}{c} a_{1} \, a_{1} ^{\prime} \\ a_{2} \, a_{2}
    ^{\prime} \end{array}
\right) \sim \MoreRep{2}{2}$ and for the product $\MoreRep{2}{2}
\times \MoreRep{2}{2}$ they read $b_{1} \, b_{2} ^{\prime} + b_{2} \, b^{\prime} _{1} \sim
\MoreRep{1}{1}$, $b_{1} \, b_{2} ^{\prime}-b_{2} \, b^{\prime} _{1}
\sim \MoreRep{1}{2}$ and $\left( \begin{array}{c} b_{2} \, b_{2} ^{\prime} \\ b_{1} \, b_{1}
    ^{\prime} \end{array}
\right) \sim \MoreRep{2}{1}$ with $\left( \begin{array}{c} b_{1} \\ b_{2} \end{array} \right) \, , \,
  \left( \begin{array}{c} b ^{\prime}_{2} \\ b ^{\prime} _{2}
      \end{array} \right) \sim \MoreRep{2}{2}$. For the mixed product
    $\MoreRep{2}{1} \times \MoreRep{2}{2}$ we find $\left(
      \begin{array}{c} a_{2} \, b_{1} \\ a_{1} \, b_{2} \end{array}
    \right) \sim \MoreRep{2}{1}$ and $\left( \begin{array}{c} a_{2} \, b_{2} \\
    a_{1} \, b_{1} \end{array} \right) \sim \MoreRep{2}{2}$ with
$a_{i}$ being the upper and lower components of $\MoreRep{2}{1}$
and $b_{i}$ of $\MoreRep{2}{2}$, respectively. 

\noindent The Clebsch Gordan coefficients for the product $\nu \times
\mu$ can be constructed from the ones given for $\mu \times \nu$ by
simply taking the transpose of these. Therefore the shown Clebsch
Gordan coefficients are sufficient for the calculation of all Yukawa
and Higgs couplings.

\noindent Finally, we display the resolution of the smallest representations of
$SO(3)$ ($SU(3)$) into irreducible ones of \Groupname{D}{5}.

\parbox{2.5in}{
\begin{eqnarray} \nonumber
&\underline{SO(3)}& \;\;\; \rightarrow \;\;\; \underline{\mbox{\Groupname{D}{5}}}\\ \nonumber
&\Rep{1}& \;\;\; \rightarrow \;\;\; \MoreRep{1}{1} \\ \nonumber
&\Rep{3}& \;\;\; \rightarrow \;\;\; \MoreRep{1}{2} + \MoreRep{2}{1}\\ \nonumber
&\Rep{5}& \;\;\; \rightarrow \;\;\; \MoreRep{1}{1} + \MoreRep{2}{1} + \MoreRep{2}{2} \\ \nonumber
&\Rep{7}& \;\;\; \rightarrow \;\;\; \MoreRep{1}{2} + \MoreRep{2}{1} +
2 \, \MoreRep{2}{2} \\ \nonumber
&\Rep{9}& \;\;\; \rightarrow \;\;\; \MoreRep{1}{1} + 2 \,
\MoreRep{2}{1} + 2 \, \MoreRep{2}{2} 
\end{eqnarray}}
\parbox{2.5in}{
\begin{eqnarray} \nonumber
&\underline{SU(3)}& \;\;\; \rightarrow \;\;\; \underline{\mbox{\Groupname{D}{5}}}\\ \nonumber
&\Rep{1}& \;\;\; \rightarrow \;\;\; \MoreRep{1}{1} \\ \nonumber
&\Rep{3}& \;\;\; \rightarrow \;\;\; \MoreRep{1}{2} + \MoreRep{2}{1} \\ \nonumber
&\Rep{6}& \;\;\; \rightarrow \;\;\; 2 \, \MoreRep{1}{1} +
\MoreRep{2}{1} + \MoreRep{2}{2}\\ \nonumber
&\Rep{8}& \;\;\; \rightarrow \;\;\; \MoreRep{1}{1} + \MoreRep{1}{2} +
2 \, \MoreRep{2}{1} + \MoreRep{2}{2} \\ \nonumber
&\Rep{10}& \;\;\; \rightarrow \;\;\; 2 \, \MoreRep{1}{2} + 2 \,
\MoreRep{2}{1} + 2 \, \MoreRep{2}{2}   
\end{eqnarray} }

\noindent One can interchange $\MoreRep{2}{1}$ with $\MoreRep{2}{2}$
to get an alternative possible embedding. These breaking sequences can
be calculated with the methods shown in \cite{breakingseq}.

\section{Tables of Numerical Examples}
\label{app:tables}

\begin{table}[h!]
\begin{center}
\scriptsize
\rule[0in]{0cm}{0cm}\begin{tabular}{|c|c|c|c|c|}
\hline
Yukawas &\rule[0.3cm]{0cm}{0cm} $\alpha ^{i}_{0}$ &  $\alpha ^{i}_{1}$ &  $\alpha ^{i}_{2}$ &
$\alpha ^{i}_{3}$ \\
\hline
$i=u$ & $-0.993413$ & $0.994834$ & $0.00047857$ & $0.000151179$ \\
$i=d$ & $-0.0169925$ & $0.0163996$ & $0.0000874833$ & $0.000127194$\\ 
$i=l$ & $-0.0107912$ & $0.00954078$ & $-0.000060939$ & $0.000056597$\\ 
$i=\nu$ & $-1.00616$ & $0.979207$ & $1.3039$ & $1.45344$\\ 
\hline 
\rule[0.2cm]{0cm}{0cm}VEVs & $\langle \chi_{1} \rangle$ & $\langle \chi_{2} \rangle$ & $\langle
 \psi_{1} \rangle$ & $\langle \psi_{2} \rangle$ \\
\hline
abs. $\left[ \GeV \right]$ & $97.3856$  & $71.386$ & $101.872$ & $68.0594$\\
phase $\left[ \mathrm{rad.} \right]$ & $-0.0076515$ & $0.0071711$ & $0.014899$ & $-$\\
\hline
\end{tabular}
\end{center}
\begin{center}
\begin{minipage}[t]{12cm}
\caption[]{ Numerical solution for Dirac neutrinos. The Yukawa
  couplings of the neutrinos have to be multiplied by
  $10^{-12.35}$ and  $\sum \limits _{i=1} ^{2} \left(|\langle \chi_{i}
  \rangle | ^{2} + |\langle \psi _{i} \rangle| ^{2} \right) = (172.02 \GeV)^{2}$. \label{SCPVnumsolDirac}}
\end{minipage}
\end{center}
\begin{center}
\scriptsize
\rule[0in]{0cm}{0cm}\begin{tabular}{|c|c|c|c|c|}
\hline
Yukawas&\rule[0.3cm]{0cm}{0cm} $\alpha ^{i}_{0}$ &  $\alpha ^{i}_{1}$ &  $\alpha ^{i}_{2}$ &
$\alpha ^{i}_{3}$ \\
\hline
$i=u$ & $-0.972907$ & $0.98535$ & $0.00044789$ & $0.000161069$ \\
$i=d$ & $-0.0166799$ & $0.016192$ & $0.0000831123$ & $0.00013383$\\ 
$i=l$ & $-0.0106418$ & $0.00937877$ & $-0.00016366$ & $0.0000211836$\\ 
$i=\nu$ & $0.83783$ & $-0.983826$ & $1.22405$ & $1.22214$\\ 
\hline 
\rule[0.2cm]{0cm}{0cm}VEVs & $\langle \chi_{1} \rangle$ & $\langle \chi_{2} \rangle$ & $\langle
 \psi_{1} \rangle$ & $\langle \psi_{2} \rangle$ \\
\hline
abs. $\left[ \GeV \right]$ & $60.385$  & $106.489$ & $56.6084$ & $110.954$\\
phase $\left[ \mathrm{rad.} \right]$ & $-0.0313569$ & $-0.0358665$ & $-0.0500026$ & $-$\\
\hline
$M_{RR}$&\multicolumn{2}{|c|}{\rule[0.4cm]{0cm}{0cm} $M_{1}= 1.878 \times 10^{14} \GeV$} &
\multicolumn{2}{|c|}{$M_{2}= 2.011 \times 10^{14} \GeV$} \\  
\hline
\end{tabular}
\end{center}
\begin{center}
\begin{minipage}[t]{12cm}
\caption[]{Numerical solution for Majorana neutrinos. The
  sum of the squares of the absolute values of the VEVs is $\approx (174.65 \GeV)^{2}$.  \label{SCPVnumsolMajo}}
\end{minipage}
\end{center}
\end{table}

\section{Experimental Data}
\label{app:experimentaldata}

\small
The masses for the quarks and charged leptons at $\mu = M_{Z}$ are \cite{massesMZ05,massesMZlept}:
\footnotesize
\[
m_u (M_{Z}) = (1.7 \pm 0.4)  \MeV \;, \;\; m_c (M_{Z}) = (0.62 \pm 0.03)
\GeV \; , \;\; m_t (M_{Z}) = (171 \pm 3) \GeV \; ,
\]
\[
m_d (M_{Z}) = (3.0 \pm 0.6) \MeV \; , \;\; m_s (M_{Z}) = (54 \pm 11)
\MeV \; , \;\; m_b (M_{Z})= (2.87 \pm 0.03) \GeV \; ,
\]
\[
m_e (M_{Z})= (0.48684727 \pm 0.00000014) \MeV \; , \;\; m_{\mu} (M_{Z})=
(102.75138 \pm 0.00033) \MeV \; ,
\]
\[
m_{\tau} (M_{Z})= 1.74669 ^{+0.00030} _{-0.00027} \GeV \; .
\]
\small
The CKM mixing angles hardly depend on the scale $\mu$ at low
energies. Therefore we take the values found in \cite{PDG} which are
measured in tree-level processes only:
\begin{eqnarray}\nonumber
&& \sin (\theta_{12}) \equiv s_{12} =0.2243 \pm 0.0016 \; , \;\; \sin
(\theta_{23}) \equiv s_{23}= 0.0413 \pm 0.0015 \; ,\\ \nonumber
&& \sin (\theta_{13}) \equiv s_{13}= 0.0037 \pm 0.0005 \; , \;\;
\delta= 1.05 \pm 0.24 \;\;\; \mbox{and} \;\;\; \mathcal{J} _{CP} =
(2.88 \pm 0.33) \, \times \, 10^{-5} \; .
\end{eqnarray}

\noindent In the neutrino sector only the two mass squared differences measured
in atmospheric and solar neutrino experiments are known \cite{nufit}:
\[
\Delsol = m_2 ^2 -m_1 ^2 = (7.9^{+ 0.6} _{- 0.6} ) \times 10^{-5} \eV
^2 \; , \;\; |\Delatm|
= |m_3 ^2 -m_1 ^2| = (2.2 ^{+ 0.7} _{-0.5} ) \times 10^{-3}  \eV^2 \; .
\]
\noindent The leptonic mixing angles are constrained: $s_{13} ^{2} \leq 0.031  \; , \;\; s_{12} ^{2}= 0.3 ^{+ 0.04} _{-0.05}$ and $s_{23} ^{2}= 0.5 ^{+ 0.14} _{-0.12}$.
All values observed in neutrino oscillations are given at $2 \,
\sigma$ level. Three further quantities connected to the neutrinos are
measurable: the sum of the neutrino masses from cosmology, $m_{\beta}$
in beta decay experiments and $|m_{ee}|$ in neutrinoless double
beta decay. The experimental bounds on these quantities
are:
\footnotesize
\begin{eqnarray}\nonumber
& & \sum \limits _{i=1} ^{3} m_{i} \leq  (0.42  \dots 1.8) \eV  \; \mbox{\cite{cosmomassbounds}} , \;\; m_{\beta} = \left( \sum \limits _{i=1}
^{3} |U_{MNS} ^{e \, i}| ^{2} \, m_{i} ^{2} \right) ^{1/2} \leq 2.2
\eV \; \mbox{\cite{betadecaybounds}} \;\; \mbox{and} \\ \nonumber & &
|m_{ee}| = |\sum \limits _{i=1} ^{3} \left( U_{MNS} ^{e \, i} \right)
^{2} \,  m_{i}| \leq 0.9 \eV \; \mbox{\cite{0vbbbounds}} \; .
\end{eqnarray}


\small

\begin{thebibliography}{99}

\bibitem{s3}   S.~Pakvasa and H.~Sugawara, Phys.\ Lett.\ B {\bf 73},
  61 (1978); H.~Harari, H.~Haut and J.~Weyers, Phys.\ Lett.\ B
  {\bf 78}, 459 (1978);   E.~Derman, Phys.\ Rev.\ D {\bf 19}, 317
  (1979);   E.~Ma, Phys.\ Rev.\ D {\bf 43}, 2761 (1991) ;  J.~Kubo, A.~Mondragon, M.~Mondragon and
  E.~Rodriguez-Jauregui, Prog.\ Theor.\ Phys.\  {\bf 109}, 795 (2003)
  [Erratum-ibid.\  {\bf 114}, 287 (2005)];   T.~Kobayashi, J.~Kubo and
  H.~Terao, Phys.\ Lett.\ B {\bf 568}, 83 (2003);   K.~Y.~Choi,
  Y.~Kajiyama, H.~M.~Lee and J.~Kubo, Phys.\ Rev.\ D {\bf 70}, 055004 (2004);  E.~Ma, Phys.\
  Rev.\ D {\bf 61}, 033012 (2000);   S.~L.~Chen, M.~Frigerio and
  E.~Ma, Phys.\ Rev.\ D {\bf 70}, 073008 (2004)
  [Erratum-ibid.\ D {\bf 70}, 079905 (2004)];  W.~Grimus and
  L.~Lavoura, JHEP {\bf 0508}, 013 (2005) ;    R.~Dermisek and
  S.~Raby, Phys.\ Lett.\ B {\bf 622}, 327 (2005); L.~J.~Hall and H.~Murayama, Phys.\
  Rev.\ Lett.\  {\bf 75}, 3985 (1995);  F.~Caravaglios and S.~Morisi, hep-ph/0510321 ; N.~Haba and K.~Yoshioka, hep-ph/0511108.

\bibitem{a4}  D.~Wyler, Phys.\ Rev.\ D {\bf 19}, 3369 (1979);
  E.~Ma and G.~Rajasekaran, Phys.\ Rev.\ D {\bf 64}, 113012 (2001);
  K.~S.~Babu, E.~Ma and J.~W.~F.~Valle, Phys.\ Lett.\ B {\bf 552}, 207
  (2003); G.~Altarelli and F.~Feruglio, Nucl.\ Phys.\ B {\bf 720},
  64 (2005);   A.~Zee, hep-ph/0508278;   G.~Altarelli and F.~Feruglio,
  hep-ph/0512103;   X.~G.~He, Y.~Y.~Keum and R.~R.~Volkas, hep-ph/0601001.

\bibitem{s4}   S.~Pakvasa and H.~Sugawara, Phys.\ Lett.\ B {\bf 82},
  105 (1979); E. ~Derman and H.-S. ~Tsao, Phys. \ Rev. \ D  {\bf
  20}, 1207 (1979); D.-G. ~Lee and R. ~N. ~Mohapatra, Phys. \ Lett. \ B {\bf 329}, 463
(1994);  R.~N.~Mohapatra, M.~K.~Parida
and G.~Rajasekaran, Phys.\ Rev.\ D {\bf 69}, 053007 (2004) ;   E.~Ma,
Phys.\ Lett.\ B {\bf 632}, 352 (2006);   C.~Hagedorn, M.~Lindner and
R.~N.~Mohapatra, arXiv:hep-ph/0602244.

\bibitem{d4} W.~Grimus and L.~Lavoura, Phys.\ Lett.\ B {\bf
    572}, 189 (2003);   W.~Grimus, A.~S.~Joshipura, S.~Kaneko,
  L.~Lavoura and M.~Tanimoto, JHEP {\bf 0407}, 078 (2004); G.~Seidl,
  hep-ph/0301044;   T.~Kobayashi, S.~Raby and R.~J.~Zhang, Nucl.\ Phys.\ B {\bf 704}, 3 (2005).

\bibitem{d2prime}  M.~Frigerio, S.~Kaneko, E.~Ma and M.~Tanimoto,
  Phys.\ Rev.\ D {\bf 71}, 011901 (2005).
 
\bibitem{d5}  E.~Ma, hep-ph/0409288.

\bibitem{dns}  C.~D.~Carone and R.~F.~Lebed, Phys.\ Rev.\ D
  {\bf 60}, 096002 (1999); S.~L.~Chen and
  E.~Ma, Phys.\ Lett.\ B {\bf 620}, 151 (2005).

\bibitem{dnprimes} P.~H.~Frampton and A.~Rasin, Phys.\ Lett.\ B {\bf
    478}, 424 (2000);   K.~S.~Babu and J.~Kubo, Phys.\ Rev.\ D {\bf
    71}, 056006 (2005);   P.~H.~Frampton and T.~W.~Kephart, Int.\ J.\
  Mod.\ Phys.\ A {\bf 10}, 4689 (1995);   P.~H.~Frampton and
  O.~C.~W.~Kong, Phys.\ Rev.\ Lett.\  {\bf 77}, 1699 (1996);
  R.~Dermisek and S.~Raby, Phys.\ Rev.\ D {\bf 62}, 015007 (2000).

\bibitem{tprime} A. ~Aranda, C. ~D. ~Carone and R. ~F. ~Lebed, Phys. \
  Lett. \ B {\bf 474}, 170 (2000); A. ~Aranda, C. ~D. ~Carone and
  R. ~F. ~Lebed, Phys.\ Rev. \ D {\bf 62}, 016009 (2000).

\bibitem{deltas}   K.~C.~Chou and Y.~L.~Wu, hep-ph/9708201;
  D.~B.~Kaplan and M.~Schmaltz, Phys.\ Rev.\ D {\bf 49}, 3741 (1994);   I.~de Medeiros Varzielas, S.~F.~King and G.~G.~Ross,
  hep-ph/0512313.

\bibitem{Lomont} J. ~S. ~Lomont, {\it Applications of Finite Groups},
  Acad. Press, 346 p. (1959).

\bibitem{JCPrel} C.~Jarlskog, Phys.\ Rev.\ Lett.\  {\bf 55}, 1039
  (1985);   H.~Fritzsch and Z.~z.~Xing, Prog.\ Part.\ Nucl.\ Phys.\
  {\bf 45}, 1 (2000).

\bibitem{masstexture}   Y.~Koide, H.~Nishiura, K.~Matsuda, T.~Kikuchi
  and T.~Fukuyama, Phys.\ Rev.\ D {\bf 66}, 093006 (2002);
  K.~Matsuda and H.~Nishiura, Phys.\ Rev.\ D {\bf 69}, 053005 (2004);
  K.~Matsuda and H.~Nishiura, Phys.\ Rev.\ D {\bf 69}, 117302 (2004).

\bibitem{mutau}  T. Fukuyama and H. Nishiura,
hep-ph/9702253; R.~N.~Mohapatra and S.~Nussinov,
Phys.\ Rev.\ D {\bf 60}, 013002 (1999); C.~S.~Lam,  Phys.\ Lett.\ B {\bf 507}, 214
(2001);  W.~Grimus and L.~Lavoura,  JHEP {\bf 0107}, 045 (2001);
P.~F.~Harrison and W.~G.~Scott,  Phys.\ Lett.\ B {\bf 547}, 219
(2002); Eur.\ Phys.\ J.\ C {\bf 28}, 123 (2003); T.~Kitabayashi
and M.~Yasue, Phys.\ Rev.\ D {\bf 67}, 015006 (2003); A.
Ghosal, hep-ph/0304090; Y.~Koide,  Phys.\ Rev.\ D {\bf 69}, 093001
(2004); R.~N.~Mohapatra,
JHEP {\bf 0410}, 027 (2004);  W.~Grimus,
A.~S.~Joshipura, S.~Kaneko, L.~Lavoura, H.~Sawanaka and M.~Tanimoto,
Nucl.\ Phys.\ B {\bf 713}, 151 (2005); T.~Kitabayashi and M.~Yasue,
hep-ph/0504212; S.~Choubey and W.~Rodejohann,
Eur.\ Phys.\ J.\ C {\bf 40}, 259 (2005);
R.~N.~Mohapatra, S.~Nasri and H.~B.~Yu, hep-ph/0502026; R. N. Mohapatra and W.
Rodejohann, hep-ph/0507319.

\bibitem{4texturezeros} Y.~F.~Zhou, arXiv:hep-ph/0309076.

\bibitem{seesaw1} P. Minkowski, Phys. Lett. {\bf B67 }, 421
(1977); M.~Gell-Mann, P.~Ramond, and R.~Slansky, \emph{Supergravity}
(P.~van Nieuwenhuizen et al. eds.), North Holland, Amsterdam, 1980,
p.~315; T.~Yanagida, in \emph{Proceedings of the Workshop on the
Unified Theory and the Baryon Number in the Universe} (O.~Sawada
and A.~Sugamoto, eds.), KEK, Tsukuba, Japan, 1979, p.~95; S.~L.
Glashow, \emph{The future of elementary particle physics}, in
\emph{Proceedings of the 1979 Carg{\`e}se Summer Institute on
Quarks and
Leptons} (M.~L{\'e}vy et al. eds.), Plenum Press, New York, 1980,
pp.~687;
R.~N. Mohapatra and G.~Senjanovi{\'c}, Phys. Rev. Lett. \textbf{44}, 912 (1980).

\bibitem{reslepto}   A.~Pilaftsis, Phys.\ Rev.\ D {\bf 56}, 5431
  (1997);   A.~Pilaftsis and T.~E.~J.~Underwood,  Nucl.\ Phys.\ B {\bf
    692}, 303 (2004); T.~Hambye, J.~March-Russell and S.~M.~West, JHEP
  {\bf 0407}, 070 (2004).

\bibitem{d3texture}   J.~Kubo, A.~Mondragon, M.~Mondragon and
  E.~Rodriguez-Jauregui, Prog.\ Theor.\ Phys.\  {\bf 109}, 795 (2003)
  [Erratum-ibid.\  {\bf 114}, 287 (2005)]. 

\bibitem{23blockmass}   G.~Altarelli and F.~Feruglio, hep-ph/0206077;
  A.~de Gouvea, Phys.\ Rev.\ D {\bf 69}, 093007 (2004).

\bibitem{fnmechanism} C.~D.~Froggatt and H.~B.~Nielsen, Nucl.\ Phys.\
  B {\bf 147}, 277 (1979).

\bibitem{cosmomassbounds}  S.~Hannestad, Nucl.\ Phys.\ Proc.\ Suppl.\
  {\bf 145}, 313 (2005) and references therein; newest WMAP results:
  \verb1http://lambda.gsfc.nasa.gov/product/map/dr2/pub_papers/1 \verb1threeyear/parameters/wmap_3yr_param.pdf1.

\bibitem{cosmomassprospect}   S.~Hannestad, Phys.\ Rev.\ D {\bf 67},
  085017 (2003);   J.~Lesgourgues, S.~Pastor and L.~Perotto, Phys.\
  Rev.\ D {\bf 70}, 045016 (2004);   M.~Kaplinghat, L.~Knox and
  Y.~S.~Song, Phys.\ Rev.\ Lett.\  {\bf 91}, 241301 (2003);
  K.~N.~Abazajian and S.~Dodelson, Phys.\ Rev.\ Lett.\  {\bf 91}, 041301 (2003).

\bibitem{theta13prospect}  F.~Ardellier {\it et al.},
  arXiv:hep-ex/0405032;   K.~Anderson {\it et al.},
  arXiv:hep-ex/0402041;   Y.~Itow {\it et al.}, arXiv:hep-ex/0106019;
  D.~Ayres {\it et al.}  [Nova Collaboration], arXiv:hep-ex/0210005;  P.~Huber, J.~Kopp, M.~Lindner, M.~Rolinec
  and W.~Winter, arXiv:hep-ph/0601266.

\bibitem{betadecaybounds}   C.~Kraus {\it et al.}, Eur.\ Phys.\ J.\ C {\bf
    40}, 447 (2005);   V.~M.~Lobashev {\it et al.}, Phys.\ Lett.\ B
  {\bf 460}, 227 (1999).

\bibitem{betadecayprospect}   A.~Osipowicz {\it et al.}  [KATRIN
  Collaboration], arXiv:hep-ex/0109033;   G.~Drexlin  [KATRIN
  Collaboration], Nucl.\ Phys.\ Proc.\ Suppl.\  {\bf 145}, 263 (2005).

\bibitem{0vbbbounds}   H.~V.~Klapdor-Kleingrothaus {\it et al.}, Eur.\
  Phys.\ J.\ A {\bf 12}, 147 (2001);  C.~E.~Aalseth {\it et al.}
  [IGEX Collaboration], Phys.\ Rev.\ D {\bf 65}, 092007 (2002). 

\bibitem{0vbbprospect} Y.~Shitov  [NEMO Collaboration],
  arXiv:nucl-ex/0405030;   S.~Capelli {\it et al.}  [CUORE
  Collaboration], arXiv:hep-ex/0505045;   R.~Ardito {\it et al.},
  arXiv:hep-ex/0501010;
  R.~Gaitskell {\it et al.}  [Majorana Collaboration],
  arXiv:nucl-ex/0311013;   I.~Abt {\it et al.}, arXiv:hep-ex/0404039;
  M.~Danilov {\it et al.}, Phys.\ Lett.\ B {\bf 480}, 12 (2000);
  H.~Ejiri, J.~Engel, R.~Hazama, P.~Krastev, N.~Kudomi and
  R.~G.~H.~Robertson, Phys.\ Rev.\ Lett.\  {\bf 85}, 2917 (2000);
  K.~Zuber, Phys.\ Lett.\ B {\bf 519}, 1 (2001);  N.~Ishihara,
  T.~Ohama and Y.~Yamada, Nucl.\ Instrum.\ Meth.\ A {\bf 373}, 325
  (1996);   S.~Yoshida {\it et al.}, Nucl.\ Phys.\ Proc.\ Suppl.\
  {\bf 138}, 214 (2005);   G.~Bellini {\it et al.}, Eur.\ Phys.\ J.\ C
  {\bf 19}, 43 (2001);   C.~Aalseth {\it et al.},
  arXiv:hep-ph/0412300.

\bibitem{d3pot}   J.~Kubo, H.~Okada and F.~Sakamaki, Phys.\ Rev.\ D
  {\bf 70}, 036007 (2004);   S.~L.~Chen, M.~Frigerio and E.~Ma, Phys.\ Rev.\ D {\bf 70}, 073008 (2004)
  [Erratum-ibid.\ D {\bf 70}, 079905 (2004)].

\bibitem{s3potreflect} S.~Pakvasa and H.~Sugawara, Phys.\ Lett.\ B
  {\bf 73}, 61 (1978).

\bibitem{S3reflectGB}   V.~Goffin, G.~Segre and H.~A.~Weldon, Phys.\
  Rev.\ D {\bf 21}, 1410 (1980);   R.~Yahalom, Phys.\ Rev.\ D {\bf
    29}, 536 (1984).

\bibitem{comingsoon} C. ~Hagedorn, {\it in preparation}.

\bibitem{lepbound}  A.~Sopczak, arXiv:hep-ph/0502002. 

\bibitem{GUTflavor} K.~Enqvist and J.~Maalampi, Nucl.\ Phys.\ B {\bf
    191}, 189 (1981); F.~Wilczek and A.~Zee, Phys.\ Rev.\ D {\bf 25},
  553 (1982); P.~Ramond, arXiv:hep-ph/9809459.

\bibitem{quantumgrav}   S.~B.~Giddings and A.~Strominger, Nucl.\
  Phys.\ B {\bf 307}, 854 (1988);   L.~M.~Krauss and F.~Wilczek,
  Phys.\ Rev.\ Lett.\  {\bf 62}, 1221 (1989).

\bibitem{A4subgroups}  G.~Altarelli and F.~Feruglio, Nucl.\ Phys.\ B
  {\bf 741}, 215 (2006); X.~G.~He, Y.~Y.~Keum and R.~R.~Volkas, arXiv:hep-ph/0601001.

\bibitem{breakingseq} M. ~Hamermesh, \emph{Group Theory and Its
    Application to Physical Problems}, Reading, Mass.: Addison-Wesley,
   509 p. (1962); J. ~Patera, R. ~T. ~Sharp, P. ~Winternitz,
  J. ~Math. ~Phys. {\bf 19}, 2362 (1978); P. ~E. ~Desmier,
  R. ~T. ~Sharp, J. ~Math. ~Phys. {\bf 20}, 74 (1979); C. ~J. ~Cummins,
  J. ~Patera, J. Math. Phys. {\bf 29}, 1736 (1988); P. ~E. ~Desmier,
  R. ~T. ~Sharp, J. ~Patera, J. ~Math. ~Phys. {\bf 23}, 1393 (1982).

\bibitem{massesMZ05} M. ~Jamin, \textit{Talk given in Bern in May
    2005}, url: \verb1http://ifae-s0.ifae.es/~jamin/1.

\bibitem{massesMZlept} H.~Fusaoka and Y.~Koide, Phys.\ Rev.\ D {\bf 57}, 3986 (1998).

\bibitem{PDG} S.~Eidelman {\it et al.}  [Particle Data Group], Phys.\
  Lett.\ B {\bf 592}, 1 (2004).

\bibitem{nufit}   M.~Maltoni, T.~Schwetz, M.~A.~Tortola and
  J.~W.~F.~Valle, New J.\ Phys.\  {\bf 6}, 122 (2004).

\end{thebibliography}
\end{document}